%% file: dist1v07.tex
\documentclass[a4paper,preprintnumbers,floatfix,superscriptaddress,showpacs,notitlepage,longbibliography]{revtex4-1}
\usepackage{graphics}
\usepackage{graphicx}
\usepackage{hyperref}
\usepackage{amsfonts}
\usepackage{amssymb}
\usepackage{pifont}
\usepackage{color}
\usepackage{amsmath,amsthm}
\usepackage{mathrsfs}
\usepackage{pgf,tikz}
\usepackage{mathrsfs}
\usetikzlibrary{arrows}

\include{comandos1}
\def\be{\begin{equation}}
\def\ee{\end{equation}}
\begin{document}
\title{On geometrical aspects of the graph approach to contextuality}
\author{Barbara Amaral}
 \affiliation{Departamento de Matem\'atica, Universidade Federal de Ouro Preto,
 Ouro Preto, MG, Brazil}
\affiliation{Departamento de F\'isica e Matem\'atica, CAP - Universidade Federal de S\~ao Jo\~ao del-Rei, 36.420-000, Ouro Branco, MG, Brazil} 
\affiliation{International Institute of Physics, Federal University of Rio Grande do Norte, 59078-970, P. O. Box 1613, Natal, Brazil}
\author{ Marcelo Terra Cunha}%\inst{2}% etc
%\thanks{\emph{Present address:} Insert the address here if needed}%                   % Do not remove
\affiliation{Departamento de Matem\'atica, Universidade Federal de Minas Gerais,
 Caixa Postal 702, 30123-970, Belo Horizonte, MG, Brazil}
\affiliation{ Departamento de Matem\'atica Aplicada, IMECC-Unicamp, 13084-970, Campinas, S\~ao Paulo, Brazil}
% % The correct dates will be entered by Springer
% %
% % Add name of the expert who has communicated your paper
% \communicated{name}

\begin{abstract}
 The connection between contextuality and graph theory has led to many developments in the field. In particular, the 
 sets of probability distributions in many contextuality scenarios
 can be described using well known convex sets  from graph theory, leading to a beautiful geometric characterization of 
 such sets. This geometry can also be explored in the definition of contextuality quantifiers based on geometric distances, which is
 important for the resource theory of contextuality, developed after the recognition of contextuality as a potential resource for quantum
 computation. In this paper we review the geometric aspects of contextuality and use it to define several quantifiers, which have the advantage of
 being applicable  to the exclusivity approach to contextuality, where previously defined quantifiers do not fit.
 \end{abstract}
\maketitle
\section{Introduction}
\label{intro}

Interesting discrepancies between  probability theories emerge in the statistics of  finite sets of measurements, divided among
several jointly measurable sets with non-empty intersection. 
One of these discrepancies is related to the 
unexpected phenomenon of \emph{quantum contextuality}, which states the impossibility of explaining
the  statistical predictions of quantum theory in terms of models where the measurement outcomes reveal pre-existent  properties that are independent  on which, or whether, other compatible measurements are jointly performed \cite{Specker60,Bell66,KS67}. This phenomenon is deeply connected to 
incompatibility of measurements  and  thus represents an exotic, intrinsically non-classical phenomenon, that 
may lead to a more fundamental understanding of the hole theory \cite{NBAASBC13,Cabello13,Cabello13c,CSW14,ATC14,Amaral14}.

Contextuality is an important advance in foundations of quantum theory, and both its theoretical and experimental aspects 
 have received much attention lately \cite{HLBBR06, KZGKGCBR09, ARBC09, LLSLRWZ11, BCAFACTP13}.    An extremely powerful framework for studying contextuality, the graph approach, was developed in Refs. \cite{CSW10,CSW14} and further  explored in Refs.
\cite{RDLTC14, AFLS15}. This framework led to  remarkable  results, consequences of the perception that the knowledge in
graph theory could be applied directly to the field of  contextuality. In particular, the 
sets of probability distributions in many contextuality scenarios
can be described using well known convex sets  from graph theory, leading to a beautiful geometric characterization of 
such sets.

In addition to  the important role of contextualtity in
foundations of physics  is
the recognition that contextuality is not just a curiosity of  quantum theory,
but a crucial \emph{resource} for quantum computing within certain special models \cite{Raussendorf13, HWVE14, DGBR14}, 
  random number certification \cite{UZZWYDDK13}, and several other information processing tasks in the specific case of space-like separated systems \cite{BCPSW13}. This has motivated the development of a \emph{resource theory of contextuality}
\cite{GHHHHJKW14, HGJKL15, ACTA17, ABM17}, in analogy to the highly developed resource-theoretic approaches to quantum nonlocality  \cite{Barrett05b, Allcock09, GWAN12, Joshi13, Vicente14, LVN14, GA15, GA17}.

Resource theories give powerful frameworks for the formal treatment of a physical property as an operational resource, adequate for its characterization, quantification, and manipulation \cite{BG15, CFS16}. They consist in the specification of three main ingredients: i) the set of objects, which specifies the physical entity that may  posses the resource; ii) a special class of transformations, called
the \emph{free operations}, that fulfill the essential requirement
of mapping every free (i.e. resourceless) object of the
theory into a free object; iii) resource quantifiers that provide a quantitative characterization of the \emph{amount} of resource  a given object contain.
In Refs. \cite{GHHHHJKW14, HGJKL15},
an abstract characterization of the axiomatic structure of
a resource theory of contextuality was developed. There, the authors define the \emph{relative entropy of contextuality}, 
a contextuality quantifier
based  on the notion of relative entropy distance, 
 also called the Kullback-Leibler divergence. The authors also mention briefly the \emph{robustness of contextuality}, a quantifier
 based on the convexity of the noncontextual set. Another quantifier based on convexity, the \emph{contextual fraction}  was introduced 
 in Refs. \cite{AB11, ADLPBC12} and further investigated  in Ref. \cite{ABM17}.
A  natural class of contextuality-free operations with a clear operational
interpretation and a explicit parametrization, the \emph{non-contextual wirings},  was  introduced in Ref. \cite{ACTA17}.

%Experiment \cite{HLBBR06, KZGKGCBR09, ARBC09, LLSLRWZ11, BCAFACTP13}.

In this contribution we review the geometric features of quantum contextuality that result from the graph approach  in both the compatibility-hypergraph approach, in which a contextuality scenario is defined by the  compatibility relations among  measurements, and the exclusivity-graph approach, in which a contextuality scenario is defined by the  exclusivity relations among  measurement events. We discuss the convex sets of probability distributions 
arising from classical, quantum and general probabilistic theories and their relations to graph invariants. In the compatibility-hypergraph approach, the noncontextual set its related to the 
\emph{cut polytope} $\mathrm{CUT}\left(\mathrm{G}\right)$ of the corresponding compatibility graph $\mathrm{G}$ and, for a special class of scenarios, to the \emph{metric polytope} $\mathrm{M}\left(\mathrm{G}\right)$ of $\mathrm{G}$. The quantum set is related to the \emph{eliptope}
$\mathcal{E}\left(\mathrm{G}\right)$ and the non-disturbing set is related to the \emph{rooted semimetric polytope} $\mathrm{RCMET}\left(\mathrm{G}\right)$. In the exclusivity-graph approach the 
classical and quantum sets of the scenario given by exclusivity graph $\mathrm{G}$ are exactly given by the stable set polytope $\mathrm{STAB}\left(\mathrm{G}\right)$  and the 
theta body $\mathrm{TH}\left(\mathrm{G}\right)$, respectively. The set of probability distributions obtained with general probabilistic theories satisfying the so called Exclusivity Principle 
is exactly given by the clique-constrained stable set polytope $\mathrm{QSTAB}\left(\mathrm{G}\right)$.
Using the geometry of these sets, we  generalize the contextuality quantifier 
introduced in Ref. \cite{GHHHHJKW14} by using  symmetric distances in the space of probabilities. 
The advantage of this approach is twofold: on one hand, we are able to link this quantifier with graph invariants and noncontextualtiy inequalities; on the other,
we can define a contextuality quantifier  in the exclusivity-graph approach, for which the quantifier presented in Ref.
\cite{GHHHHJKW14} does not fit. We also list some of the important features of the convexity based contextuality quantifiers.

\section{The Compatibility-Hypergraph Approach}
\label{sectioncontextcompat}

\begin{defi}
 A \emph{compatibility scenario} is given by a triple $\Upsilon :=\left(\mathcal{M}, \mathcal{C}, O \right)$, where $O$ is a finite
  set, $\mathcal{M}$ is a finite set of random variables in  $\left(O, \mathcal{P}\left(O\right)\right)$, and 
  $\mathcal{C}$ is a family of subsets  of $\mathcal{M}$ such that
\begin{enumerate}
 \item $\cup_{C \in \mathcal{C}} C =\mathcal{M}$;
\item   $C,C' \in \mathcal{C}$ and $C \subseteq C'$
implies $C = C'$. \label {antichain}
\end{enumerate}
The elements $C \in \mathcal{C}$ are called (maximal) \emph{contexts} and the
set $\mathcal{C}$ is called the \emph{compatibility cover} of the scenario.
\end{defi}

The random variables in  $\mathcal{M}$ represent measurements with possible outcomes $O$
in a physical system and the sets in $\mathcal{C}$ encode the compatibility relations among the elements of $\mathcal{M}$, i. e.,
each  $C \in \mathcal{C}$ consists of a maximal set of  measurements that can be jointly performed.

\begin{defi}
The \emph{compatibility hypergraph} of  $\left(\mathcal{M}, \mathcal{C}, O \right)$ is the hypergraph $\mathrm{H} = \left(\mathcal{M}, \mathcal{C}\right)$
whose vertex-set is $\mathcal{M}$ and  edge-set is  $\mathcal{C}$.
The \emph{compatibility graph} of the scenario is  the 2-section of $\mathrm{H}$, that is, the graph $\mathrm{G}$ 
 with vertex-set $\mathcal{M}$ and edge-set
 \be E\left(\mathrm{G}\right)= \left\{\left(M_i, M_j\right) \left| \exists \ e   \in 
 E\left(\mathrm{H} \right) \  ; \  M_i, M_j \in e\right.\right\}.\ee

\end{defi}

The fact that a set of measurements is pairwise compatible does not necessarily imply that this set is jointly compatible. Hence, in general, the compatibility hypergraph is more subtle than its 2-section.  
 Nonetheless, if only quantum contextuality with projective measurements is to be discussed,
 pairwise compatibility implies joint compatibility.  The  maximal contexts in this compatibility cover
 correspond to the maximal cliques of the compatibility graph. 

For a given context $C \in \mathcal{C}$,  the set of possible outcomes for a joint measurement of the elements of $C$ is the Cartesian product of $|C|$ copies of $O$, denoted by $O^{C}$. %This set can  be identified with the set of functions
%\be \lambda: C \longrightarrow O.\ee
%Each function $\lambda \in O^C$ is called a \emph{section over} $C$.
When  the measurements in $C$ are  jointly performed, a set of outcomes in $O^C$
will be observed. This individual run of the experiment will be called an \emph{measurement event}. 

\begin{defi}
A \emph{behaviour} $\mathrm{B}$ for the scenario $\left(\mathcal{M}, \mathcal{C}, O \right)$ is a family of probability distributions over $O^C$, one for each context
$C \in \mathcal{C}$, that is, 
\be \mathrm{B} = \left\{p_C:  O^C\rightarrow [0,1] \left|\sum_{s\in O^C} p_C(s)=1, C \in \mathcal{C}\right.\right\}.\ee
\end{defi}

For each $C$, $p_C(s)$ gives the probability of obtaining outcomes $s$ in a joint measurement of the elements of $C$.
It will be convenient to associate each behaviour  to a vector in $\mathbb{R}^N,$ where 
$N=\displaystyle{\sum_{C \in \mathcal{C}}\left|O^C\right|}$, that we will  denote by $P_\mathrm{B}$.
If we have $\mathcal{C}=\left\{C_1, C_2, \ldots, C_n\right\}$ and for each $C_i$ we have
$O^{C_i}= \left\{s_i^1, s_i ^2, \ldots , s_i^{m_i}\right\}$, we define
\be
 P_\mathrm{B}=\left[\begin{array}{ccccccc}
             p_{C_1}\left(s_1^1\right)&\ldots&p_{C_1}\left(s_1^{m_1}\right)&  \ldots&p_{C_n}\left(s_n^1\right)&\ldots&p_{C_n}\left(s_n^{m_n}\right)
            \end{array}\right]. \ee
%This association is discussed in more detail in reference \cite{AQBTC13}. 

%For a given compatibility cover, the set of possible behaviors is a polytope with
%$\displaystyle{\prod_{C \in \mathcal{C}}\left|O^{C}\right|}$ vertices. Each vertex corresponds to 
%probability one for one of the outcomes
%$s \in O^C$ for each context $C \in \mathcal{C}$.
%All other behaviors are convex combinations of these vertices.

Let $C=\{M_1, \ldots, M_m\}$ be a context in $\mathcal{C}$. Each element of $O^C$ is a string $s= \left(a_1, \ldots, a_m\right)$ with
 $m$ elements of $O$. For each $U \subset C$, there is a natural restriction
\begin{eqnarray}
 r^C_U:O^C &\rightarrow& O^U \nonumber\\
s=\left(a_i\right)_{M_i \in C} &\mapsto &s|_U=\left(a_i\right)_{M_i \in U}.
\label{eqrestriction}
\end{eqnarray}
%This operation corresponds to dropping the elements in the string $s$ that do not correspond to measurements in $U$.

Given a  probability distribution in $C \in \mathcal{C}$ we can also naturally define  marginal distributions for each  $U\subset C$:
\begin{eqnarray}
 p^C_{U}\ :\ O^{U}&\rightarrow &[0,1] \nonumber\\
p^C_{U}(s)&=& \sum_{s' \in O^C; r^C_U(s') = s} p_C(s').
\end{eqnarray}
The superscript $C$ in $p^C_U$ is necessary  because, without further restrictions, the marginals may depend on the context $C$.

\begin{defi}
\label{definondisturbance}
The \emph{non-disturbance} set $\mathcal{X}\left(\Upsilon\right)$ is the set of behaviors such that for any two intersecting  contexts $C$ and $C'$
 the consistency relation \be p^C_{C\cap C'}= p^{C'}_{C\cap C'} \ee holds. 
%A probability distribution $p \in \mathcal{X}\left(\Gamma\right)$ is
%called an \emph{empirical model}.
\end{defi}

The non-disturbance set is a polytope, since it is defined by a finite number of linear inequalities and equalities. 

We
ask now if it is possible to extend the distributions $p_C$ to larger sets containing $C$ in a consistent way. The naive ultimate 
goal would be to define a distribution on
the set $O^\mathcal{M}$, which specifies assignment of outcomes to all measurements, in a way that the restrictions
 yield the probabilities specified by the behaviour on all
 contexts in $\mathcal{C}$. A more subtle and adequate question is to decide when it is possible to 
achieve this goal. This question was first studied by Fine in Ref.  \cite{Fine82}, for the 
restricted case of Bell scenarios,  and 
generalized by Brandenburger
 and Abramsky  in Ref.  \cite{AB11}.
As it happens in many branches of mathematics, the notion of contextuality is
deeply connected to the possibility of extending elements of $O^C$ to global sections in $O^{\mathcal{M}}$. 

\begin{defi}
A \emph{global section} for $\mathcal{M}$ is a probability distribution $p_{\mathcal{M}}: O^\mathcal{M} \rightarrow  [0,1]$. A \emph{global section for a behaviour} 
$\mathrm{B} \in \mathcal{X}\left(\Upsilon\right)$
is a global section for $\mathcal{M}$
such that the restriction of $p_{\mathcal{M}}$ to each context $C \in \mathcal{C}$ is equal to $p_C$.  The behaviors with global section
are called \emph{noncontextual}.
\end{defi}

Behaviors with global section  are deeply connected with \emph{noncontextual completions} of quantum theory, also known as 
\emph{noncontextual hidden variable models} \cite{AB11, AT17}.

\subsection{Classical Realizations and Non-contextuality}
 \begin{defi}
A \emph{classical realization} for the scenario $\Upsilon=\left(\mathcal{M}, \mathcal{C}, O \right)$
is given by a  probability space $(\Omega,\Sigma, \mu )$, where $\Omega$ is a sample space, 
$\Sigma$ a $\sigma -$algebra and $\mu$ a probability measure in $\Sigma$, and for each $M \in \mathcal{M}$ a partition
of $\Omega$ into $|O|$ disjoint subsets $A_j^M \in \Sigma, \ \ j \in O$.
% we can say that a distribution is non-contextual if for each $i\in V$ there is a random variable $R_i : \Omega \rightarrow O$
% and $p(a_1, \ldots, a_n|M_1, \ldots, M_n)= \mu\left(R_i=a_i\right).$}
For each context $C= \{M_1, \ldots, M_m\}$, the probability of the outcome $s=\left(a_1, \ldots, a_m\right)$ for a joint
measurement of the elements of $C$ is 
\be p_C(s)=\mu\left(\bigcap_{k=1}^m A_{a_k}^{M_k}\right). \label{eq:classical_realization}\ee 
 The behaviors that can be obtained in this form are called \emph{classical} or 
 \emph{non-contextual behaviors}.
The set of all classical behaviors will be denoted by
$\mathcal{C}\left(\Upsilon\right)$. 
\end{defi}

The set 
$\mathcal{C}\left(\Upsilon\right)$ is a polytope with $\left|O^{\mathcal{M}}\right|$ vertices.
If a behaviour $\mathrm{B}$ is classical, we have that
\be p_U^C\left(s|_U\right)=\mu\left(\bigcap_{k \left| M_k \in U\right.} A_{a_k}^{M_k}\right)\ee
is independent of the context $C$, and hence $\mathrm{B} \in \mathcal{X}\left(\Upsilon\right).$

\begin{prop}
\label{propclassicalnoncontextual}
 A behaviour has a global section if and only if it is classical.
\end{prop}

For a proof of this result, see  Refs. \cite{Amaral14, AT17}.
%\dem{
%Once a classical realization is given, the construction of the global section is guaranteed by 
%the fact that the intersection of a finite number of sets in a $\sigma$-algebra also belongs to the $\sigma$-algebra. Conversely, given the 
%global section $p_{\mathcal{M}}$, we use $\Omega = O^ {\mathcal{M}}$, $\Sigma = \mathcal{P}\left(\Omega\right)$ and $\mu(s)=p_{\mathcal{M}}(s)$ as a classical realization.
%This result shows that the existence of a global section is an alternative definition for classicality, making clear its topological character.
%The facets of this polytope are defined by
%linear inequalities
%}

\subsection{Quantum Realizations}
\label{subsec:quantum_beh_comp}

\begin{defi}
A \emph{quantum realization} 
for the scenario $\Upsilon=\left(\mathcal{M}, \mathcal{C}, O \right)$ is given by a Hilbert space $\mathcal{H}$, for each $M \in \mathcal{M}$ a
partition of the identity operator acting in $\mathcal{H}$ into $\left|O\right|$  projectors $P_j^{M}$, $j \in O$, 
  and a density matrix $\rho$ acting on $\mathcal{H}$. For a given context $C=\{M_1, \ldots, M_m\} \in \mathcal{C}$,
the compatibility condition demands the existence of a basis for $\mathcal{H}$ in
which all  $P_j^{M_i}$ are diagonal, or, equivalently, \be \De{P_j^{M_i},P_l^{M_k}} = 0, \forall i,j,k,l. \ee
The probability of the outcome $s=\left(a_1, \ldots, a_m\right)$ for a joint measurement of $C$  is 
\be p_C(s)=\tr\left(\prod_{k=1}^m P_{a_k}^{M_k}\rho\right).\ee
 The behaviors that can be written in this form  are called \emph{quantum  behaviors}. The set of all quantum behaviors will
 be denoted by $\mathcal{Q}(\Upsilon)$.
\end{defi}

 It is a known fact that  $\mathcal{Q}(\Upsilon)$  is convex and $\mathcal{C}(\Upsilon) \subset \mathcal{Q}(\Upsilon)$. It is  not a polytope in general. Notice
that the Hilbert space is not fixed and the set  $\mathcal{Q}(\Upsilon)$  contains realizations in all dimensions.

 \section{Contextuality Quantifiers}
 
 \label{sectionprop}

\begin{defi}
A resource theory for contextuality is defined by a set $\mathcal{F}$ of linear operations $\mathcal{T}: \mathcal{X}\left(\Upsilon\right)\rightarrow 
\mathcal{X}\left(\Upsilon\right)$ such that \be \mathcal{T}\left[\mathcal{C}\left(\Upsilon\right)\right] \subset \mathcal{C}\left(\Upsilon\right).\ee
A function $X: \mathcal{X}\left(\Upsilon\right)\rightarrow \mathbb{R}$ is a \emph{contextuality monotone} for this resource theory of contextuality if
 \be X\left[\mathcal{T}\left(B\right)\right] \leq X\left(B\right)\ee
 for every $\mathcal{T} \in \mathcal{F}$.
 \end{defi}

Besides monotonicity under free operations, other properties of a monotone $X$ are also desirable \cite{HGJKL15, ABM17}:

\begin{enumerate}

\item  \emph{Faithfullness:} For all $B \in \mathcal{C}(\Upsilon)$, $X\left(B\right)=0$.

\item  \emph{Preservation under reversible operations:} If $\mathcal{T} \in \mathcal{F}$ is reversible, then
\be X\left(\mathcal{T}\left(B\right)\right) = X\left(B\right).\ee

 %The set of allowed transformations is presented in reference \cite{}.
%General transformations can be written in the block form 
%$$\mathcal{T}= \left[\begin{array}{ccc}
 %                \alpha_{11} T_{11} & \cdots& \alpha_{1N} T_{1N}\\
 %                \vdots&\ddots& \vdots\\
  %               \alpha_{N1}T_{N1}&\ldots &\alpha_{NN}T_{NN}          
   %            \end{array}\right]$$
%where $T_{ij}$ is a stochastic matrix taking a distribution on context $j$ to a distribution in context $i$ and
%$\sum_j \alpha_{ij}=1$. 

%Since a stochastic transformation decreases the norm of a vector, monotonicity is valid for $D$.

\item  \emph{Additivity:} We consider two kinds of additivity. First we consider a scenario $\Upsilon$ such that its compatibility hypergraph $\mathrm{H}=\mathrm{H}_1 \& \mathrm{H}_2$ consists of two
connected components $\mathrm{H}_1$ 
and
$\mathrm{H}_2$. The behaviors for $\Upsilon$ are formed by the list of probabilities for the scenario given by $\mathrm{H}_1$ 
followed by the list of  probabilities
for the scenario given by $\mathrm{H}_2$. It follows that   any behavior $B=B_1 \& B_2$  in $\mathrm{H}$ is the juxtaposition of a 
behavior $B_1$  for $\mathrm{H}_1$ and a behavior $B_2$ for $\mathrm{H}_2$ and hence the quantifier $X$ should be such that 
\be
X\left(B_1 \& B_2\right)\leq X\left(B_1\right)+X\left(B_2\right).
\ee
One may also require that equality holds.

Another kind of operation we can apply to two scenarios is considering that all measurements in $\mathrm{H}_1$ are compatible with all
measurements in
$\mathrm{H}_2$, but with the restriction that they should be independent. 
This implies that a behavior for $\mathrm{H}$
is the tensor product of one behavior for $\mathrm{H}_1$ with a behavior for $\mathrm{H}_2$. For this kind of operation,
subadditivity of $X$ should hold. 
%\begin{eqnarray*}
%X\left(P \otimes P_{\mathcal{C}'}\right)\leq D\left(P \otimes P_{\mathcal{C}'}, Q_{\mathcal{C}} \otimes Q_{\mathcal{C}'}\right) &=& \sum_{k,l} |p_kp'_l - q_kq'_l|\\
%&=&
%\sum_{k,l} |p_kp'_l - p_kq'_l +p_kq'_l- q_kq'_l|\\
%&\leq &\sum_{k,l} p_k|p'_l - q'_l| + q'_l|p_k - q_k|\\
%&=&D\left(P , Q_{\mathcal{C}} \right)+D\left( P_{\mathcal{C}'}, Q_{\mathcal{C}'}\right)\\
%&=&X\left(P\right)+X\left(P_{\mathcal{C'}}\right).
%\end{eqnarray*}
\be
X\left(B_1 \otimes B_2\right)\leq X\left(B_1\right)+X\left(B_2\right).
\ee

\item  \emph{Convexity:} If a behavior can be written as $B=\sum_i \pi_i B^i$, where $\pi_i \in [0,1]$ and each $B^i$ is a behavior  for the same scenario, then 
\be X(B) \leq \sum_i \pi_i X\left(B^i\right).\ee

\item  \emph{Continuity:} $X\left(B\right)$ should be a continuos function of $B$.

\end{enumerate}
 
In what follows we exhibit a number of monotones for different resource theories of contextuality and
 list which of the  properties above they satisfy.
 
 \subsection{Entropic Contextuality Quantifiers}
 \label{sectionentropicquant}

 In Ref. \cite{GHHHHJKW14} the authors introduce  contextuality quantifiers based on two distinct approaches. 
 The first one uses a communication game
 to grasp the phenomenon of contextuality in a quantitative manner. The second just postulates a measure, called 
 \emph{relative entropy of contextuality}, analogous to similar non-locality quantifiers defined in Ref. \cite{DGG05}. The two approaches are equivalent, since 
 the quantifier that emerges in the communication game equals the relative entropy of contextuality.
 
 \subsubsection{The ``which context'' game}
 
Three players, Alice, Bob and Charlie, pre-agree on some fixed behavior $B$ in a given contextuality scenario $\Upsilon=\left(\mathcal{M}, \mathcal{C}, O \right)$. 
 The goal of Alice is to communicate
  a context $C$ to Bob, through the hands of the adversary Charlie, whose goal is to 
 stop the communication of $C$ to Bob. 
 To this end,  Alice chooses $C$ according to some probability
 distribution $\pi(C)$ and sends it  to Charlie, which creates a global section $p_{\mathcal{M}}^C$ in 
 such a way that it is compatible with
 the distribution $p_C$ given by  $B$ and then sends it to Bob. Bob's goal is to guess the context $C$ sent by Alice.
 
 If $B$ is noncontextual, the existence of a global distribution $p_{\mathcal{M}}$ for $B$ guarantees that Charlie can choose $p_{\mathcal{M}}^C=p_{\mathcal{M}}$ in such a 
 way that it marginalizes to $p_{C'}$ for every
 context $C'\in \mathcal{C}$,
 and Bob will have no information about $C$. On the other hand, if $B$ is contextual, there is at least one context $C'$ for which 
 $p_{\mathcal{M}}^C$ and $p_{C'}$ do not agree, and Bob will  have a better guess for $C$.

If, after this procedure, we denote the  amount of correlations between Alice and Bob by $I_{\pi}(B)$
 %\be I_{p}(B)=\min_{P^J_I} I\left(\sum_J p(J)\ket{J}\bra{J} \otimes P_I^J\right).\ee
 and maximize it over all probability distributions $\pi$ defined over the set of contexts $\mathcal{C}$, we obtain the \emph{mutual information of contextuality}
 \be I_{max}\left(B\right)=\sup_{\pi} I_{\pi}\left(B\right).\ee
 The value of $I_{max}$ quantifies how much correlations Alice and Bob have after the procedure, and can also be seen as a quantifier of
 how much the a priori behavior $B$ is noncontextual (see reference \cite{GHHHHJKW14} for details). 
 
 \subsubsection{Relative entropy of contextuality}
 
  In Ref. \cite{GHHHHJKW14}, the authors also introduce two measures of contextuality based directly on the notion of relative entropy distance, 
 also called the Kullback-Leibler divergence. Given two probability distributions $p$ and $q$ in a sample space $\Omega$, the 
 Kullback-Leiber divergence between $p$ and $q$ 
 \be D_{\mathrm{KL}}(p\|q) = \sum_{i\in \Omega} p(i) \, \log\frac{p(i)}{q(i)} \ee 
 is a measure of  the difference between the two probability distributions $p$ and $q$.
 
 \begin{defi}
 The \emph{Relative Entropy of 
 Contextuality} of a behavior $B$ is defined as
 \begin{equation}\label{HorDist2}
E_{max}\left(B \right) =   \min _{B^{NC} \in \mathcal{C}\left(\Upsilon\right) }\ \ \max _{\pi}\ \ 
\sum _{C\in \mathcal{C}}\ \pi(C)\ D_{\mathrm{KL}}\left(p_C  \middle\| p^{NC}_C \right),
\end{equation}
where the minimum is taken over all noncontextual behaviors $B^{NC}=\left\{ p^{NC}_C \right\}$ and the maximum is taken over all
 probability distributions $\pi$ defined on the set of contexts $\mathcal{C}$.
The \emph{Uniform Relative Entropy of Contextuality} of $B$
is defined as
\begin{equation}\label{HorDist3}
E_{u}\left(B \right) = \frac{1}{N} \min _{B^{NC}\in \mathcal{C}\left(\Upsilon\right) } \ \ \sum _{C\in \mathcal{C}}\ D_{\mathrm{KL}}\left(p_C  \middle\| p^{NC}_C \right),
\end{equation}
 where $N=\left|\mathcal{C}\right|$ is the number of contexts in $\mathcal{C}$ and, once more, the minimum is taken over all noncontextual behaviors $B^{NC}=\left\{ p^{NC}_C \right\}$.
 \end{defi}
 
 Both can be interpreted  as a ``distance'' of the bahavior $B$ to the set of noncontextual models $\mathcal{C}\left(\Upsilon\right).$
 
 Interestingly, the Relative Entropy of 
 Contextuality is equal to the quantifier based on the ``which context'' game \cite{GHHHHJKW14}:
\be I_{max}=E_{max}.\ee

In reference \cite{ACTA17} it is shown that $E_{max}$ is a monotone under \emph{noncontextual wirings}. The quantity $E_{u}$, however, is not a monotone under the complete class of 
noncontextual wirings, as shown in Ref. \cite{GA17} for the special class of 
Bell scenarios. Nonetheless, it is a monotone under a broad class of such operations. More specifically, it is monotone under
post-processing operations and under a subclass of pre-processing operations.

\begin{teo}
\label{teo:prop_entropic}
\begin{enumerate}
\item $E_{max}$ is a contextuality monotone for the resource theory of contextuality defined by  noncontextual wirings;
\item $E_u$ is a contextuality monotone for the resource theory of contextuality defined by post-processing operations and a subclass of pre-processing operations;
\item $E_{max}$ and $E_u$ are faithful, additive, convex, continuous, and preserved under relabellings of inputs and outputs.
\end{enumerate}
\end{teo}

 The proof of this result can be found in Appendix \ref{app:entropic}.
 
 \subsection{Geometric Contextualtiy Quantifiers}
 \label{sectiongeometric}

We now introduce contextuality monotones  based on geometric distances, in contrast 
with the previous defined quantifiers which are based on entropic distances.
Let $D$ be any distance defined in real vector spaces  $\mathbb{R}^K$. 
 The first quantifier we propose is based on the distance of the vector $P_B$ to the set of vectors obtained with noncontextual behaviors:

\begin{defi}
The $D$-\emph{contextuality distance} of a behavior $B$ is defined as
 \be \mathcal{D}\left(B\right)=\min _{B^{NC}\in \mathcal{C}\left(\Upsilon\right) }  D\left(P_B , P_{B^{NC}} \right).
 \label{eqdefdist}\ee
 \end{defi}

We can also calculate the distance between the behaviors $B$ and $B^{NC}$ for each context $C$ and then averaging over the contexts. When the choice of context is uniform, we have:

\begin{defi}
The $D$-\emph{uniform contextuality distance} of a behavior $B$ is defined as
\be \mathcal{D}_u\left(B\right)=\frac{1}{N}\min_{B^{NC}\in \mathcal{C}\left(\Upsilon\right) } \sum_{C\in \mathcal{C}} D\left(p_C ,p^{NC}_C \right), \label{eqdefdist2}\ee
where $N=\left|\mathcal{C}\right|$ is the number of contexts in $\mathcal{C}$.
\end{defi}

If we allow a non-uniform choice of context, the natural way of quantifying contextuality will be:

\begin{defi}
The $D$-\emph{max contextuality distance} of a behavior $B$ is defined as
\be \mathcal{D}_{max}\left(B\right)= \min _{B^{NC}\in \mathcal{C}\left(\Upsilon\right) } \max_{\pi} \sum _{C\in \mathcal{C}}\ \pi(C)\ D\left(p_C, p^{NC}_C \right),  \label{eqdefdist3}\ee
where the minimum is taken over all noncontextual behaviors $B^{NC}=\left\{ p^{NC}_C \right\}$ and the maximum is taken over all
over all probability distributions $\pi$ defined over the set of contexts $\mathcal{C}$.
\end{defi}

The quantifiers $\mathcal{D}_{u}$ and $\mathcal{D}_{max}$ are just special cases of $\mathcal{D}$, since we obtain them using a proper choice of distance in Eq.
\eqref{eqdefdist}. Nevertheless, we  stress out these definitions because of their physical meaning and special mathematical properties (see Thm. \ref{teo_dist} below). 
%\be D_1(p)=\frac{1}{N}\inf_Q d(p,q), \ q \in \mathcal{NC}(\Gamma). \label{eqdefdist}\ee

%\be D_2(p)=\frac{1}{N}\inf_Q \sum_C d(p_C,q_C), \ q \in \mathcal{NC}(\Gamma). \label{eqdefdist2}\ee

%\be D_3(p)=\sup_{P(C)}\inf_Q \sum_C P(C)d(p_C,q_C), \ q \in \mathcal{NC}(\Gamma),  \ 0 \leq P(C) \leq 1, \sum_CP(C)=1. \label{eqdefdist3}\ee

%The measures $D_2$ and $D_3$ can be seen as special cases of $D_1$.

Calculating exact values for these quantifiers is not an easy computational problem in general. For example, if $D$ is the distance obtained with the $\ell_1$ and 
$\ell_2$ norms, although the 
minimization can be done efficiently in the number of vertices of the set of noncontextual behaviors using linear
and quadratic programming, respectively,  the number of vertices grows enormously if the compatibility
graph gets more complicated, which makes the problem intractable in general for a large number of vertices.
Nonetheless, we can calculate these distances for some interesting examples (see Sub. \ref{sub:n_cycle} and, for the special class of
Bell scenarios, Ref. \cite{BAC17}).

The properties satisfied by the quantities defined in Eqs. \eqref{eqdefdist}, \eqref{eqdefdist2} and \eqref{eqdefdist3} will depend on the distance $D$ used in the definition. We focus our
attention on distances defined by $\ell_P$ norms:

\begin{teo}
\label{teo_dist}
\begin{enumerate}
\item $\mathcal{D}_{max}$ is a contextuality monotone for the resource theory of contextuality defined by the noncontextual wiring operations;
\item $\mathcal{D}_u$ is a contextuality monotone for the resource theory of contextuality defined by post-processing operations and a subclass of pre-processing operations;
\item $\mathcal{D}$, $\mathcal{D}_{u}$ and $\mathcal{D}_{max}$ are faithful, additive, convex, continuous, and preserved under relabellings of inputs and outputs.
\end{enumerate}
\end{teo}

This result is proven in Appendix \ref{ap:dist}. It shows that while $\mathcal{D}_{max}$ is a proper 
contextuality monotone under the entire  class of noncontextual wirings,
$D$ and $D_u$ are more suitable when the set of allowed free operations preserves the scenario under consideration.

\section{Contextual Fraction}
\label{sec:cf}

A contextuality quantifier based on the intuitive notion of what \emph{fraction} of a given behavior admits a noncontextual description
was introduced in Refs. \cite{AB11, ADLPBC12}. Several properties of this quantifier were further discussed in Ref. \cite{ABM17}.

\begin{defi}
The \emph{contextual fraction} of a behavior $B$ is defined as
\be
\label{eq:cont_frac}
\mathcal{CF}\left(B\right)= \min \left\{\lambda \left|B=  \lambda B' + \left(1-\lambda\right)B^{NC}\right.\right\},
\ee
where $B^{NC}$ is an arbitrary noncontextual behavior.
\end{defi}

\begin{teo}
 The contextual fraction is a monotone under all linear operations that preserve the classical set $\mathcal{C}\left(\Upsilon\right)$.
\end{teo}

\dem{Let $\mathcal{T}$ be a linear operation over the set of behaviors such that 
\be
\mathcal{T}\left(\mathcal{C}\left(\Upsilon\right)\right) \subset \mathcal{C}\left(\Upsilon\right).
\ee
Given a behavior $B$, let $B= \lambda B' + \left(1-\lambda\right)B^{NC}$ be the decomposition of $B$ achieving the 
minimum in Eq. \eqref{eq:cont_frac}, that is, $\mathcal{CF}(B)=\lambda$. Then
\begin{eqnarray}
 \mathcal{T}\left(B\right) &=&\mathcal{T}\left(\lambda B' + \left(1-\lambda\right)B^{NC}\right)\\
 &=&\lambda\mathcal{T}\left( B' \right) + \left(1-\lambda\right)\mathcal{T}\left(B^{NC}\right).
\end{eqnarray}
Since $\mathcal{T}\left(B^{NC}\right)$ is a noncontextual behavior, we conclude that
\be
\mathcal{CF}\left(\mathcal{T}\left(B\right)\right) \leq \lambda= \mathcal{CF}\left(B\right).
\ee
\hfill\qed\vspace{1ex}
}

Moreover, the contextual fraction also satisfies:

\begin{prop}
 \label{teo:cf}
 \begin{enumerate}
  \item The contextual fraction is faithful, convex and continuous;
  \item $\mathcal{CF}\left(B_1 \& B_2\right) \leq \max_i \mathcal{CF}\left(B_i\right)$;
  \item $\mathcal{CF}\left(B_1 \otimes  B_2\right) \leq \mathcal{CF}\left(B_1\right) + \mathcal{CF}\left(B_2\right) - \mathcal{CF}\left(B_1\right)\mathcal{CF}\left(B_2\right)$;
 \item The contextual fraction can be calculated via linear programming.
 \end{enumerate}

\end{prop}

The proof of these results can be found in Ref. \cite{ABM17}.

\section{Robustness of Contextuality}
\label{sec:rob}

The Robustness of contextuality is a quantifier based on the intuitive notion of how much noncontextual  \emph{noise} 
 a given behavior can sustain before becoming noncontextual  \cite{HGJKL15}. 
 
 \begin{defi}
The \emph{robustness} of a behavior $B$ is defined as
\be
\label{eq:rob}
\mathcal{R}\left(B\right)= \min \left\{\lambda \left|  \left(1-\lambda\right) B + \lambda B^{NC} \in \mathcal{C}\left(\Upsilon\right)\right.\right\},
\ee
where $B^{NC}$ is an arbitrary noncontextual behavior.
\end{defi}

\begin{teo}
 The robustness of contextuality is a monotone under all linear operations that preserve the classical set $\mathcal{C}\left(\Upsilon\right)$.
\end{teo}

\dem{Let $\mathcal{T}$ be a linear operation over the set of behaviors such that 
\be
\mathcal{T}\left(\mathcal{C}\left(\Upsilon\right)\right) \subset \mathcal{C}\left(\Upsilon\right).
\ee
Given a behavior $B$, let $\left(1-\lambda\right)  B + \lambda B^{NC}$ be the decomposition of $B$ achieving the 
minimum in Eq. \eqref{eq:rob}, that is, $\mathcal{R}(B)=\lambda$. Then
\be
 \mathcal{T}\left(\left(1-\lambda\right)  B + \lambda B^{NC}\right)\\
 =\left(1-\lambda\right)\mathcal{T}\left( B \right) + \lambda\mathcal{T}\left(B^{NC}\right) \in \mathcal{C}\left(\Upsilon\right).
\ee
Since $\mathcal{T}\left(B^{NC}\right)$ is a noncontextual behavior, we conclude that
\be
\mathcal{R}\left(\mathcal{T}\left(B\right)\right) \leq \lambda= \mathcal{R}\left(B\right).
\ee
\hfill\qed\vspace{1ex}
}

Moreover, the robustness of contextuality also satisfies:

\begin{teo}
 \begin{enumerate}
  \item The robustness of contextuality is faithful, convex and continuous;
  \item $\mathcal{R}\left(B_1 \& B_2\right) \leq \max_i \mathcal{R}\left(B_i\right)$;
  \item $\mathcal{R}\left(B_1 \otimes  B_2\right) \leq \mathcal{R}\left(B_1\right) + \mathcal{R}\left(B_2\right) - \mathcal{R}\left(B_1\right)\mathcal{R}\left(B_2\right)$;
 \item The contextual fraction can be calculated via linear programming.
 \end{enumerate}
\label{teo:rob}
\end{teo}

The proof of this result can be found in Appendix \ref{sub:rob}.

\section{The Geometry of scenarios with $H=G$ and $|O|=2$}
\label{sectionn22}

If $\Upsilon$ is a scenario in which every context consists of at most two measurements, the compatibility hypergraph $H$ is equal to the 
compatibility graph  $G$.
If each measurement has two outcomes, labeled from now on $\pm 1$, both nondisturbing  and  noncontextual sets can be equivalently described in  different ways
that lead to   familiar polytopes from graph theory. Adapting the definitions of Sub. \ref{sectiongeometric} to this description we can define
other contextuality monotones for this  family of contextuality scenarios.

\subsection{Description of the  nondisturbing, quantum and noncontextual behaviors}

In this type of scenario, the nondisturbing set $\mathcal{X}\left(\Upsilon\right)$ is a subset of $\mathbb{R}^{4|E|}$.
Given a context $\{M_i,M_j\} \in \mathcal{C}$ we denote by $p_{ij}(ab)$ the probability of obtaining outcome $a$ for measurement $M_i$ and
outcome $b$ for measurement $M_j$. We denote by $p_{i}(a)=\sum_b p_{ij}(ab)$ the  marginal probability for measurement $M_i$ and similar 
for measurement $M_j$.

The conditions imposed on the behavior $B$ allows us to determine all its entries knowing only $p_{ij}(-1-1)$ and $p_{i}(-1)$. In fact, we can define
\begin{eqnarray}
\phi:\mathbb{R}^{4\left|E\left(G\right)\right|} &\longrightarrow &\mathbb{R}^{\left|V\left(G\right)\right| + \left|E\left(G\right)\right|}\\
B&\longmapsto &q=\left(q_i,q_{kj}\right)_{i\in V\left(G\right); (k,j)\in E\left(G\right)}
\end{eqnarray}
where $q$ is such that $q_i=p_{i}(-1)$ and $q_{ij}=p_{ij}(-1-1)$. To recover $B$ from $q$ just notice that
\begin{eqnarray}
 p_{ij}(-1+1) &= &q_i - q_{ij} \\
  p_{ij}(+1-1) &=& q_j - q_{ij} \\
   p_{ij}(+1+1) &=&1 - q_i - q_j + q_{ij}.
   \end{eqnarray}
   
   It happens that the image of all nondisturbing behaviors for this scenario under the action of transformation $\phi$ is equal to a well known 
   convex polytope from graph theory, the \emph{correlation polytope} of $G$.
   
   \begin{defi}
   Given $S \subset V(G)$, we define the correlation vector $v(S)\in \mathbb{R}^{|V(G)|+|E(G)|}$

\begin{eqnarray*}
v(S)_i&=&\left\{\begin{array}{cc}
1& \mbox{if} \  i \in S;\\
0& \mbox{otherwise}.
\end{array}\right. \\
v(S)_{ij}&=&\left\{\begin{array}{cc}
1& \mbox{if} \ i, j \in S;\\
0& \mbox{otherwise}.
\end{array}\right.
\end{eqnarray*}

The correlation polytope $\mathrm{COR}(G)$ is the convex hull of all correlation vectors.
\end{defi}

 Notice that the correlation vectors correspond to the image of 
the extremal behaviors in $\mathcal{C}(G)$ under the action of $\phi$, which proves the following result:

\begin{teo} If $\Upsilon$ is a scenario for which $H=G$, then $\phi(\mathcal{C}\left(\Upsilon\right))=\mathrm{COR}(G)$.
\end{teo}

%\textcolor{blue}{Referência: Geometry of cuts and metrics, pag 53.}
The image of the non-disturbance polytope is also a well known polytope from graph theory.

\begin{defi}
The \emph{rooted correlation semimetric polytope} $\mathrm{RCMET}\left(G\right)$ of a graph $G$ is the set of vectors $q=\left(q_i, q_{jk}\right)\in \mathbb{R}^{\left|V\left(G\right)\right| + \left|E\left(G\right)\right|}$   such that
\begin{eqnarray}
q_{ij} &\geq &0,\\
 q_i - q_{ij} &\geq &0,\\
  1 - q_i -q_j + q_{ij} &\geq & 0.
  \end{eqnarray}
\end{defi}

\begin{prop} If $\Upsilon$ is a scenario for which $H=G$, then $\phi(\mathcal{X}\left(\Upsilon\right))=\mathrm{RCMET}(G)$.
\end{prop}

For a proof of this result, see Ref. \cite{AT17}.
%\begin{proof}

%\end{proof}

\subsection{The Cut Polytope}

\begin{defi}
Given a graph $G$ and  $c \in \{-1, 1\}^{\left|V\left(G\right)\right|}$, the cut vector of $G$ defined by $c$ is the vector  $x(c) \in \mathbb{R}^{\left|E\left(G\right)\right|}$
such that
\be x(c)_{ij}=c_ic_j.\ee
The \emph{cut polytope} of $G$, $\mathrm{CUT}^{\pm 1} (G)$, is the convex hull of all cut vectors of $G$.
\end{defi}

There exists a relation between the polytopes $\mathrm{CUT}$ and $\mathrm{COR}$.

\begin{defi}
The \emph{suspension graph} $\nabla G$ of  $G$ is the graph with vertex-set $V\left(G\right) \cup\left\{e\right\}$ and edge-set
 $E\left(G\right)\cup \left\{(e,i), i \in V\left(G\right)\right\}$. 
 \end{defi}
 
 Intuitively, $\nabla G$ is the graph obtained from $G$ by adding an extra vertex and connecting it to all vertices of $G$.

\begin{prop}
$\mathrm{CUT}^{\pm 1}\left(\nabla G\right)=\psi \left(\mathrm{COR}(G)\right),$
in which
\begin{eqnarray}
\psi: \mathbb{R}^{\left|V\left(G\right)\right|+\left|E\left(G\right)\right|} &\longrightarrow& \mathbb{R}^{\left|V\left(G\right)\right|+\left|E\left(G\right)\right|}\\
q&\longmapsto &x
\end{eqnarray}
and the coordinates of $x$ are given by
\be \begin{array}{cc}
x_{ij}=1-2q_{i}-2q_j + 4q_{ij},& (i,j) \in E\left(G\right)\\
x_{ei}=1-2q_i, & i \in V\left(G\right)
\end{array}.\ee
\end{prop}

%\textcolor{blue}{Referência: Geometry of cuts and metrics, pag 55.
%On the Relationship between Convex Bodies
%Related to Correlation Experiments with Dichotomic Observables, seção 3.}
For a proof of this result, see Refs. \cite{DL97,AIT06}.

We can interpret  $x$ in terms of expectation values of the measurements $M_i$ in the scenario:
\begin{eqnarray}
x_{ij}&=&\langle M_iM_j\rangle =p_{ij}(11) + p_{ij}(-1-1) - p_{ij}(-11) - p_{ij}(1-1)\\
x_{ei}&=&\langle M_i\rangle =p_{i}(1) - p_{i}(-1)
\end{eqnarray}

We can also define the cut polytope using the outcome values $0$ e $1$ instead of $\pm 1$.

\begin{defi}
Given a graph $G$ and  $c \in \{0, 1\}^{\left|V\left(G\right)\right|}$, the $01$-cut vector of $G$ defined by $c$ is the vector  $y(c) \in \mathbb{R}^{\left|E\left(G\right)\right|}$
such that
\be y(c)_{ij}=c_ic_j.\ee
The $01$-\emph{cut polytope} of $G$, $\mathrm{CUT}^{01} (G)$, is the convex hull of all $01$-cut vectors of $G$.
\end{defi}

The two definitions $\mathrm{CUT}^{\pm 1}$ and $\mathrm{CUT}^{01}$ are related by a bijective linear map
\begin{eqnarray}
\alpha: \mathrm{CUT}^{\pm 1}(G)& \longrightarrow &\mathrm{CUT}^{01}(G)\\
x&\longmapsto &y\\
y_{ij}&=&1-2x_{ij}.
\end{eqnarray}

All these polytopes are  hard to characterize for general scenarios. This happens because the number of extremal points grows enormously with 
the number of vertices in $G$. Hence we look for connections between these polytopes and other simpler polytopes, even if this connections is only valid 
for a restricted class of graphs.
Following this idea, for some graphs it is possible to relate $\mathrm{CUT}^{01}(G)$  with the so called \emph{metric polytope} of $G$,
denoted by $\mathrm{MET}(G)$.

\begin{prop}
$\mathrm{CUT}^{01}(G)=\mathrm{MET}(G)$ if, and only if, $G$ has no $K_5$ minor.
\label{teocut=met}
\end{prop}

This result is extremely useful since $\mathrm{MET}(G)$ is easily characterized by the following result:

\begin{prop}
\label{metineq}
Given $F \subset E\left(G\right)$ and $y \in \mathbb{R}^{\left|E\left(G\right)\right|}$, let 
\be y(F)=\sum_{(i,j) \in F} y_{ij}.\ee The following are true for $\mathrm{MET}(G)$:
\begin{enumerate}
\item $\mathrm{MET}(G)=\{y \in \mathbb{R}^{\left|E\left(G\right)\right|} | \ y_{ij} \leq 1, \  y(F)-y(C\setminus F) \leq |F|-1, \ C \ \mbox{cycle of} \ G, F \subset C, |F| \ \mbox{odd}\};$
\item The inequality $y(F)-y(C\setminus F) \leq |F|-1$ defines a facet of $\mathrm{MET}(G)$ if, and only if, $C$ is a chordless cycle;
\item The inequality $y_{ij} \leq 1$ defines a facet of $\mathrm{MET}(G)$ if, and only if, the edge $(i,j)$ does not belong to a triangle of $G$.
\end{enumerate}
\end{prop}

Props. \ref{teocut=met} and  \ref{metineq} can be used to find all facets of $\mathrm{CUT}^{01}(G)$ if $G$ has no $K_5$-minor. In this case, the facets are
defined by the so called \emph{n-cycle inequalities}: 
\be y(F)-y(C\setminus F) \leq |F|-1, \ C \ \mbox{cycle of} \ G, F \subset C, |F| \ \mbox{odd}.\ee
We can use these inequalities and the map $\alpha$ to find the facet-defining inequalities of $\mathrm{CUT}^{\pm 1}(G)$, 
if $G$ has no $K_5$-minor, which are given by
\be x(F)-x(C\setminus F) \leq |C|-2, \ C \ \mbox{cycle of} \ G, F \subset C, |F| \ \mbox{odd}.\ee
This is the same set of inequalities found for the special case $G=C_n$ in Ref. \cite{AQBTC13}.

 A similar result is valid for $\mathrm{RCMET}(G)$.

%\textcolor{blue}{Referência: Geometry of cuts and metrics, pg 432, teorema 27.3.3.}
%\vspace{1em}

\begin{prop}The image of $\mathrm{RCMET}(G)$ under $\psi$ is the \emph{rooted semimetric politope} of $\nabla G$, $\mathrm{RMET}(\nabla G)$.
\label{teoRCMET=RMET}
\end{prop}

The proofs of Thms. \ref{teocut=met}, \ref{metineq}, \ref{teoRCMET=RMET} and many other properties of these polytopes
can be found in Refs. \cite{DL97, AT17}. As a corollary, we have:

\begin{cor}
If $\Upsilon$ is a scenario for which $H=G$, then $\psi \circ \phi(\mathcal{C}\left(\Upsilon\right))=\mathrm{CUT}^{\pm 1}\left(\nabla G\right)$,
$\alpha \circ \psi \circ \phi(\mathcal{C}\left(\Upsilon\right))=\mathrm{CUT}^{01}\left(\nabla G\right)$, and 
$\psi \circ \phi(\mathcal{X}\left(\Upsilon\right))=\mathrm{RMET}\left(\nabla G\right)$.
\end{cor}

\subsection{Correlation functions}

To describe completely the sets $\mathcal{X}\left(\Upsilon\right)$, $\mathcal{Q}\left(\Upsilon\right)$ and $\mathcal{C}\left(\Upsilon\right)$ for scenarios with at most two measurements per context using the convex bodies 
defined in the previous sections,  we have to use 
vectors in  $\mathbb{R}^{\left|V\left(G\right)\right|+\left|E\left(G\right)\right|}$. In some situations  it might be useful to consider a projection of these vectors in $\mathbb{R}^{\left|E\left(G\right)\right|}$ obtained
by eliminating the coordinates relative to the edges $(e,i)$:
\begin{eqnarray}
\Pi: \mathbb{R}^{\left|V\left(G\right)\right|+\left|E\left(G\right)\right|} & \longrightarrow & \mathbb{R}^{\left|E\left(G\right)\right|}\\
x=(x_{ei}, x_{jk})_{i \in \left|V\left(G\right)\right|; (j,k) \in \left|E\left(G\right)\right|} & \longmapsto & (x_{jk})_{(j,k) \in 
\left|E\left(G\right)\right|}
\end{eqnarray}

The vectors in  $\Pi\left(\mathrm{RMET}(G)\right)$ are called \emph{correlation vectors}.

\begin{prop} Given a graph $G$ the following are true:
\begin{enumerate}
\item $\Pi\left(\mathrm{RMET}(G)\right)=[-1,1]^{\left|E\left(G\right)\right|};$
\item $\Pi\left(\mathrm{CUT}^{\pm 1}(\nabla G)\right)=\mathrm{CUT}^{\pm 1}( G).$
\end{enumerate}
\end{prop}

See Ref. \cite{DL97} for a proof. Notice that the knowledge of the correlation functions is not enough to fully recover the behavior, 
since we are loosing the information on the marginals when we apply the projection $\Pi$. Nonetheless, these vectors may be useful for two reasons: first, they provide a simpler description of the behaviors, which give some information in scenarios where $\nabla G$ is too complicated to deal with; second, although correlation vectors
do not give full information about the behavior, 
they can be enough to decide whether the corresponding  behaviors are  \emph{ contextual or not}. For example, if $G=C_n$ the knowledge of $\Pi(x)$ is enough to 
decide membership in $\mathcal{C}\left(\Upsilon\right)$, as shown in Ref. \cite{AQBTC13}.

The set $\Pi\left(\mathcal{Q}\left(\Upsilon\right)\right)$ is much harder to characterize.

\subsection{The eliptope and the set of quantum behaviors}

For Bell scenarios with two parties, one with  $n$ measurements at her disposal  and the other with $m$ measurements at her disposal, the corresponding graph is the complete bipartite graph $K_{m,n}$. In this particular type of scenario, the set  $\Pi(\mathcal{Q})$ is related to the \emph{eliptope} of the graph $G$.

\begin{prop}
The following are true
\begin{enumerate}
\item $z=(\langle M_iM_j\rangle) \in \Pi(\mathcal{Q})$;
\item There are vectors $u_i, v_j \in R^d, 1 \leq i \leq m, 1 \leq j \leq n, d \leq m + n$, such that
\be  z_{ij} = \braket{u_i}{v_j} .\ee \label{eliptopo}
\end{enumerate}
\label{teo:eliptope_q}
\end{prop}

For a proof of this result, see Ref. \cite{AII06}.

\begin{defi}
The eliptope $ \mathcal{E}(G)$ of a graph $G$ is set of vectors $x \in \mathbb{R}^{\left|E\left(G\right)\right|}$ such that for each
$i \in V\left(G\right)$  exists a unit vector $  u_i \ \in \ \mathbb{R}^{\left|V\left(G\right)\right|}$ such that 
\be x_{ij}=\braket{u_i}{u_j}.\ee
\end{defi}

 With this definition, Thm. \ref{teo:eliptope_q} states that the set of quantum correlation vectors in a bipartite Bell scenario is the eliptope of $K_{m,n}$.

The natural question is whether Thm. \ref{teo:eliptope_q} is also valid for general contextuality scenarios, that is, we want to know if given any
 graph $G$, the equality $\Pi\left(\mathcal{Q}\left(\Upsilon\right)\right)=\mathcal{E}(G)$ holds. The inclusion $\Pi\left(\mathcal{Q}\left(\Upsilon\right)\right) \subset \mathcal{E}(G)$ is always true.

\begin{teo}
$\Pi\left(\mathcal{Q}\left(\Upsilon\right)\right)\subset \mathcal{E}(G)$.
\end{teo}

See Appendix \ref{approp} for a proof of this result.

%\textcolor{blue}{Referência: Quantum Correlations: From Bell inequalities to Tsirelsons
%theorem, David Avis.}

%\vspace{1em}

For some graphs the inclusion $\mathcal{E}(G) \subset \Pi\left(\mathcal{Q}\left(\Upsilon\right)\right)$ does not hold. This is the case for
the  $n$-cycle $C_n$ for any odd $n$.  This is shown by the fact that the violation of the $n$-cycle inequalities for some points in the eliptope 
can be larger than the  maximum violation obtained with quantum models.

\begin{teo}
\label{propmaxel}
There is a point $z \in \mathcal{E}\left(C_n\right)$ for wich  
\be \sum_{i=0}^{n-2} z_{ii+1} - z_{0n-1} = n\cos\left(\frac{\pi}{n}\right).\ee
\end{teo}

This point is explained in detail in appendix \ref{approp}. Its existence proves that, in general, $\Pi(\mathcal{Q}(G))\neq \mathcal{E}(G)$. For any $n$ odd, Thm. \ref{propmaxel} shows that there is an element 
for which $\sum_{i=0}^{n-2} z_{ii+1} - z_{0n-1} = n\cos\left(\frac{\pi}{n}\right)$, while the quantum maximum for this same quantity is 
\be \frac{3n\cos\left(\frac{\pi}{n}\right) -n}{1+\cos\left(\frac{\pi}{n}\right)}\leq n\cos\left(\frac{\pi}{n}\right).\ee

Another family of graphs for which $\mathcal{E}(G)$ is different from the quantum set are the complete graphs $K_n$. In this case, all
measurements are compatible and hence the quantum set is equal to the classical set, a polytope. On the other hand, 
$\mathcal{E}(G)$ is a polytope if, and only if, $G$ is a forest \cite{DL97}, and in this case $\mathrm{CUT}^{\pm 1} (G)=\mathcal{E}(G)=\Pi\left(\mathrm{RMET}(G)\right)=[-1,1]^{|E|}.$  
%The eliptope $K_3=C_3$ is shown in Figure \ref{figeliptopeK3}. 

For the  $n$-cycles with  $n$ even, $\mathcal{E}\left(C_n\right)= \Pi(\mathcal{Q}(G))$. This is a consequence of the fact that in this case $C_n$ is
a subgraph of the complete bipartite graph $K_{n/2,n/2}$ and the eliptope of  $C_n$ is a projection of the eliptope of $K_{n/2,n/2}$.

%\begin{figure}[h]
%	\centering
%		\includegraphics[scale=0.3]{politopos.png}
%		\caption{The eliptope of the graph $\mathrm{K}_3$.}
%		\label{figeliptopeK3}
%\end{figure}

\section{Contextuality monotones for  scenarios with $H=G$ and $|O|=2$}
\label{sectionusinggeometry}

%\textcolor{blue}{Referência: Geometry of cuts and metrics, corolário 31.3.15, pg 522. Figura: Complexity of the positive semidefinite matrix
%completion problem with a rank constraint}

In the previous Section, we have shown that we can use different polytopes to characterize the set of behaviors in the particular case where
contexts have at most two measurements and each measurement has two outcomes.  In any representation we choose, the non-disturbance,
quantum and non-contextual sets are convex sets in $\mathbb{R}^{\left|V\left(G\right)\right|+\left|E\left(G\right)\right|}$ with full dimension and 
we can use a distance $D$ defined  in $\mathbb{R}^{\left|V\left(G\right)\right|+\left|E\left(G\right)\right|} $ to quantify contextuality.

In Sub. \ref{sectiongeometric} we defined 
contextuality monotones using distances defined in real vectors spaces when the behaviors are describe by the vector $p_{ij}(ab)$. Using the same idea,
we can define contextualty quantifiers

\begin{defi} When the behaviors are described by the
the vectors $q = \phi(p) \in \phi\left[\mathcal{X}\left(\Upsilon\right)\right]=\mathrm{RCMET}\left(G\right)$, we can define the contextuality quantifiers
\begin{eqnarray} \mathcal{D}^{\phi}\left(q\right)=\frac{1}{\left|E\left(G\right)\right|}\min _{q^{NC} \in \mathrm{COR}\left(G\right) } D\left(q , q^{NC} \right),\label{eqdefdist_1_cor}\\
\mathcal{D}_u^{\phi}\left(q\right)=\frac{1}{\left|E\left(G\right)\right|}\min _{q^{NC} \in \mathrm{COR}\left(G\right) }\sum_{(i,j) \in E\left(G\right)}
 D\left(q_{[ij]} , q_{[ij]}^{NC} \right),\label{eqdefdist_2_cor}\\
 \mathcal{D}_{max}^{\phi}\left(q\right)=\min _{q^{NC} \in \mathrm{COR}\left(G\right) }\max_{(i,j) \in E\left(G\right)}
 D\left(q_{[ij]} , q_{[ij]}^{NC} \right),\label{eqdefdist_3_cor}
\end{eqnarray}
 where $q_{[ij]}=\left(q_i,q_j, q_{ij}\right)$.
\end{defi}
 
\begin{defi} When the behaviors are described by 
the vectors $x \in \psi \circ \phi\left[\mathcal{X}\left(\Upsilon\right)\right]=\mathrm{RMET}\left(G\right)$, we can define 
\begin{eqnarray} \mathcal{D}^{\psi}\left(x\right)=\frac{1}{\left|E\left(G\right)\right|}\min _{x^{NC} \in \mathrm{CUT}^ {\pm 1}\left( \nabla G\right) } D\left(x , x^{NC} \right),\label{eqdefdist_1_cutpm1}\\
\mathcal{D}_u^{\psi}\left(x\right)=\frac{1}{\left|E\left(G\right)\right|}\min _{x^{NC} \in \mathrm{CUT}^ {\pm 1}\left(\nabla G\right) }\sum_{(i,j) \in E\left(G\right)}
 D\left(x_{[ij]} , x_{[ij]}^{NC} \right),\label{eqdefdist_2_cutpm1}\\
 \mathcal{D}_{max}^{\psi}\left(x\right)=\min _{x^{NC} \in \mathrm{CUT}^ {\pm 1}\left(\nabla G\right) }\max_{(i,j) \in E\left(G\right)}
 D\left(x_{[ij]} , x_{[ij]}^{NC} \right),\label{eqdefdist_3_cutpm1}
\end{eqnarray}
where $x_{[ij]}= \left(x_i, x_j, x_{ij}\right)$.
\label{defi:dist_cut_pm1}
\end{defi}

\begin{defi}
 When the behaviors are described by 
the vectors $y \in \alpha \circ \phi \circ \psi\left[\mathcal{X}\left(\Upsilon\right)\right]$, we define
\begin{eqnarray} \mathcal{D}^{\alpha}\left(y\right)=\frac{1}{\left|E\left(G\right)\right|}\min _{y^{NC} \in \mathrm{CUT}^ {01}\left( \nabla G\right) } D\left(y , y^{NC} \right),\label{eqdefdist_1_cut01}\\
\mathcal{D}_u^{\alpha}\left(y\right)=\frac{1}{\left|E\left(G\right)\right|}\min _{y^{NC} \in \mathrm{CUT}^ {01}\left(\nabla G\right) }\sum_{(i,j) \in E\left(G\right)}
 D\left(y_{[ij]} , y_{[ij]}^{NC} \right),\label{eqdefdist_2_cut01}\\
 \mathcal{D}_{max}^{\alpha}\left(x\right)=\min _{y^{NC} \in \mathrm{CUT}^ {01}\left(\nabla G\right) }\max_{(i,j) \in E\left(G\right)}
 D\left(y_{[ij]} , y_{[ij]}^{NC} \right),\label{eqdefdist_3_cut01}
\end{eqnarray}
where $y_{[ij]}= \left(y_i, y_j, y_{ij}\right)$.
\end{defi}

In any case, we have a contextuality quantifier satisfying all the properties listed in Sub. \ref{sectiongeometric}. As we already mention, it is not trivial to compute these
quantities for general graphs due to the complexity of the polytopes $\mathrm{COR}\left(G\right)$, $\mathrm{CUT}^ {\pm 1}\left( \nabla G\right)$ and 
$\mathrm{CUT}^ {01}\left( \nabla G\right)$. 
Nevertheless, all of them have full 
 dimension $\left|V\left(G\right)\right| + \left|E\left(G\right)\right|$, and its  facets  are hyperplanes with maximum dimension,
which makes the computation of these quantifiers possible in some particular scenarios. In Sub. \ref{sub:n_cycle}, we show an analytical expression for $\mathcal{D}^{\psi}\left(x\right)$ in the $n$-cycle scenario when $D$ is defined by a $\ell_p$ norm.
%In any case we can at least easily calculate 
%upper bounds.

We can also define the contextual fraction and the robustness of contextuality in these descriptions. 
If we use 
the vectors $q \in \mathrm{RCMET}\left(G\right)$, we can define 
\begin{eqnarray}
 \mathcal{F}^{\psi}\left(q\right)&=& \min \left\{\lambda \left|q=  \lambda q' + \left(1-\lambda\right)q^{NC}\right.\right\},\\
 \mathcal{R}^{\psi}\left(q\right)&=& \min \left\{\lambda \left| \left(1-\lambda\right) q + \lambda q^{NC} \in \mathrm{COR}\left(G\right)\right.\right\},
\end{eqnarray}
where $q^{NC} \in \mathrm{COR}\left(G\right)$ is arbitrary.
Analogously, if we use 
the vectors $x \in \mathrm{RMET}\left(\nabla G\right)$, we can define 
\begin{eqnarray}
 \mathcal{F}^{\phi}\left(x\right)&=& \min \left\{\lambda \left|x=  \lambda x' + \left(1-\lambda\right)x^{NC}\right.\right\},\\
 \mathcal{R}^{\phi}\left(x\right)&=& \min \left\{\lambda \left| \left(1-\lambda\right) x + \lambda x^{NC} \in \mathrm{CUT}^{\pm 1}\left(\nabla G\right)\right.\right\},
\end{eqnarray}
where $x^{NC} \in \mathrm{CUT}^{\pm 1}\left(G\right)$ is arbitrary.
If we use 
the vectors $y \in \alpha \left[ \mathrm{RMET}\left(\nabla G\right)\right]$, we can define 
\begin{eqnarray}
 \mathcal{F}^{\alpha}\left(y\right)&=& \min \left\{\lambda \left|y=  \lambda y' + \left(1-\lambda\right)y^{NC}\right.\right\},\\
 \mathcal{R}^{\alpha}\left(y\right)&=& \min \left\{\lambda \left| \left(1-\lambda\right) y + \lambda y^{NC} \in \mathrm{CUT}^{01}\left(\nabla G\right)\right.\right\},
\end{eqnarray}
where $y^{NC} \in \mathrm{CUT}^{01}\left(G\right)$ is arbitrary.
In any case, we obtain a quantifier satisfying the properties listed in Secs. \ref{sec:cf} and \ref{sec:rob}, which can be computed via linear programming.

\subsection{The $n$-cycle}
\label{sub:n_cycle}

When the compatibility graph of the scenario is the $n$-cycle $G=C_n$, the hypothesis of  Thm. \ref{metineq} are satisfied, and hence
 the facets of the cut polytope $\mathrm{CUT}^{\pm 1}\left(\nabla C_n\right)$ are defined by   the $n$-cycle inequalities
\be x(F)-x(C_n\setminus F) \leq n-2, \  F \subset C_n, |F| \ \mbox{odd}.\ee
 In this case, $\mathcal{D}$ can be easily computed. In fact, each contextual behavior violates 
only one of these inequalities, and hence the distance
of such a point to the set of non-contextual behaviors is equal to the distance of this distribution
to the hyperplane defining the facet.

Given $x \notin \mathrm{CUT}^{\pm 1}(\nabla C_n)$,  suppose $x(F)-x(C_n\setminus F) \leq n-2$ is the inequality which $x$ violates. 
If the distance  $D$ is define by any $\ell_p$-norm 
in $\mathbb{R}^{\left|V\left(G\right)\right|+\left|E\left(G\right)\right|}=\mathbb{R}^ {2n} $,
 the distance from  $x$  to $\mathrm{CUT}^{\pm 1}(\nabla G)$
is given by
\be   \frac{x(F)-x(C_n\setminus F) -n+2}{\sqrt[q]{n}} \ee
where $q \in \mathbb{N}$ is such that 
\be \frac{1}{p}+\frac{1}{q}=1.\ee
Hence, we have 
\be \mathcal{D}_{p}^{\psi}(x)= \frac{x(F)-x(C_n\setminus F) -n+2}{n\sqrt[q]{n}}  .\ee
%The factor $n=|E|$ in the denominator assures that the measures $X_1$ are asymptotic continuous. 
In  particular, for the  $\ell_2$ norm  we have
\be \mathcal{D}_2^{\psi}(x)= \frac{x(F)-x(C_n\setminus F) -n+2}{n\sqrt[2]{n}}  .\ee 
for  the $\ell_1$  norm we have \be \mathcal{D}_1^{\psi}(x)=\frac{x(F)-x(C_n\setminus F) -n+2}{n}  \ee
 and for  the maximum norm we have
\be \mathcal{D}_{\infty}^{\psi}(x)=\frac{x(F)-x(C\setminus F) -n+2}{n^2},\ee
where $\mathcal{D}_p^{\psi}$ denotes the $\mathcal{D}^{\psi}$ when computed with the $\ell_p$ norm.

%The proofs of the above results are sketched in  \ref{approp}.
The same argument can be used whenever the contextual behavior violates only one facet-defining inequality for 
$\mathrm{CUT}^{\pm 1}\left(\nabla G\right)$. To calculate $\mathcal{D}_p^{\psi}$ it suffices to identify which inequality the behavior
 violates and calculate the 
distance from the point corresponding to the behavior to the facet defined by the inequality. Unfortunately, since $\mathrm{CUT}^{\pm 1}\left(\nabla G\right)$ has an intricate 
structure,  in the general case the behavior can violate more than one facet-defining inequality.
For example, in the $(3,3,2,2)$  Bell scenario we can find a behavior
which violates  the CHSH inequality and the $I_{3322}$ inequality, both facet-defining. The detailed discussion of this example can be found in 
Appendix \ref{app:3322}.

\subsection{Connection to graph invariants}

To any scenario we can associate   a  graph $\mathcal{G}$ whose vertices  are the measurement events 
and the edges link exclusive events \cite{CSW14, Amaral14, AT17}.
We say that two events are exclusive if in both of them a same measurement was performed and for this 
measurement different outcome were obtained. We will refer to $\mathcal{G}$ as the
\emph{exclusivity graph} of the experiment \cite{CSW14, AT17}. The \emph{exclusivity graph} $\mathcal{G}_I$ of  a non-contextuality inequality is the induced subgraph of $\mathcal{G}$ defined by
the vertices that correspond to events appearing in the inequality.

It happens that if a noncontextuality inequality is written in terms of the probabilities 
$p_C(s)$, the classical and quantum maxima for this inequality are related to the graph invariants of  $\mathcal{G}_I$ \cite{CSW14}.
We can then use graph invariants to 
calculate the distances $\mathcal{D}$ defined above, or at least obtain upper bounds in the worst case scenario.

\begin{prop}[Cabello, Severini and Winter, 2010]
Given the sum 
\be \sum_i \gamma_i p_{C_i}  \left(s_i\right),\label{eq:ineq} \ee
 the maximum value attained with classical behaviors  is the vertex-weighted independence number $\alpha\left(\mathcal{G}_I, \gamma\right)$ 
 and the
  the maximum value attained with quantum behaviors is upper bounded by the vertex-weighted Lov\'asz  number $\vartheta\left(\mathcal{G}_I, \gamma\right)$ of the exclusivity
 graph $\mathcal{G}_I$ of the inequality with vertex weights given by the coefficients $\gamma_i$ of the sum \eqref{eq:ineq}.
 \label{teoqbound}
\end{prop}

We also consider probability distributions obtained when we use generalized probability theories, 
but  satisfying the following principle:

\begin{prin}[The Exclusivity Principle]
\label{defiexclusivityprinciple}
Given a set  $\{e_k\}$  of pairwise exclusive events, the corresponding probabilities $p_k$  satisfy the 
 following equation:
\begin{equation}
 \sum_{k} p_k \leq 1.
\end{equation}
\end{prin}

From now on, we refer to the Exclusivity principle simply as the \emph{E-principle}.
From the graph theoretical point of view, this restriction is equivalent to impose the condition that
whenever the set of vertices $\{v_k\}$ is a clique
in $\mathcal{G}_I$, the sum of the corresponding probabilities $p_k$ can
not exceed one.
%Specker pointed out that, in
%quantum theory, pairwise joint measurability of a set $\mathcal{M}$
%of observables implies joint measurability of $\mathcal{M}$, while in
%other theories this implication does necessarily hold \cite{Specker60}. This property is known as the \emph{Specker principle}.
%Later, Specker conjectured that this is \emph{the fundamental
%theorem} of quantum theory \cite{Specker09}. The E-principle is a consequence of the Specker principle, as shown in reference 
%\cite{NBAASBC13}. 
%The E principle can be used to explain why (some) models outside the quantum set are forbidden. 
A detailed discussion of the E-principle and its consequences can be found in Refs. \cite{NBAASBC13,CSW14,Cabello13,Cabello13c,ATC14,
Amaral14, AT17}.

The  maximum for models satisfying the E-principle is also  related to the graph $\mathcal{G}_I$.

\begin{prop}[Cabello, Severini and Winter, 2010]
\label{teoebound}
 The  maximum value for the sum  \eqref{eq:ineq}
  attained with behaviors satisfying the E-principle is equal to the vertex-weighted fractional  packing number  $\alpha^*\left(\mathcal{G}_I, \gamma\right)$ of the exclusivity
 graph $\mathcal{G}_I$ of the inequality.
\end{prop}

\subsubsection{The $n$-cycle scenario}

Since 
\begin{eqnarray}
\left\langle M_iM_j \right\rangle &=& 2\left(p_{ij}(11)+p_{ij}(-1-1)\right)-1 \nonumber\\
-\left\langle M_iM_j \right\rangle &=& 2\left(p_{ij}(p(1-1)+p_{ij}(-11)\right)-1,
\end{eqnarray}
 there are
 $2n$ events in each noncontextuality inequality for the $n$-cycle scenario. If $n$ is odd, the corresponding 
 exclusivity graph is 
 the \emph{prism graph} of order $n$, $Y_n$,  and if $n$ is even, the exclusivity graph is 
 the \emph{M\"obius ladder} of order $2n$, $M_{2n}$ \cite{AQBTC13}. %The first four
 %of these graphs are depicted in figure \ref{figncycle}. 
 
 %\begin{figure}
 %\centering
 % \includegraphics[scale=0.4]{ncycle.png}
%\caption{Exclusivity graphs for the $n$-cycle inequalities for $n=3,4,5,6$.}
%\label{figncycle}
% \end{figure} 
 
 The observation that 
 \begin{eqnarray}
  \vartheta\left(Y_n\right)&=&\frac{3n\cos\left(\frac{\pi}{n}\right)-n}{1 + \cos\left(\frac{\pi}{n}\right)},\\
 \vartheta\left(M_{2n}\right)&=& n\cos\left(\frac{\pi}{n}\right)\end{eqnarray} and Thm. \ref{teoqbound} were used by the authors in 
 Ref. \cite{AQBTC13} to 
 find the 
 quantum maximum violation of the $n$-cycle inequalities, which in this case coincides with the Lov\'asz number of the exclusivity graph. The classical bound is 
 equal to $n$ for $n$ even and $n-1$ for $n$ odd, while the E-principle bound is equal to $2n$ for every $n$.
 This allows us to compute the maximum value of $\mathcal{D}_p^{\psi}$ is this scenario, which gives us the following result:
 
 \begin{teo}
 The maximum value of $\mathcal{D}_p^{\psi}$ for the $n$-cicle scenario attainable with quantum behaviors is
 \be  \frac{\vartheta\left( Y_n\right)-\alpha\left( Y_n\right)}{n\sqrt[p]{n}}\ee
 for $n$ odd and 
  \be  \frac{\vartheta\left( M_{2n}\right)-\alpha\left( M_{2n}\right)}{n\sqrt[p]{n}}\ee
  for $n$ even.
  The maximum value of $\mathcal{D}_p^{\psi}$ for the $n$-cicle scenario attainable with E-principle behaviors is
 \be  \frac{\alpha^*\left( Y_n\right)-\alpha\left( Y_n\right)}{n\sqrt[p]{n}}\ee
 for $n$ odd and 
  \be  \frac{\alpha^*\left( M_{2n}\right)-\alpha\left( M_{2n}\right)}{n\sqrt[p]{n}}\ee
  for $n$ even.
 \end{teo}

\section{The Exclusivity-Graph Approach}
\label{sectionexclusivity}

In the  \emph{exclusivity-graph approach}, we start with  a graph $G$ which encodes the exclusiveness relations among the different 
measurement events in the scenario \cite{CSW14, Amaral14, AT17}.

\begin{defi}
A \emph{contextuality scenario} in the exclusivity-graph approach is define by a graph $G$ whose 
vertices $i \in V\left(G\right)$ are associated to measurement events $\Pi_i $  in some probabilistic model such $\Pi_i$ and $\Pi_j$ are exclusive whenever 
$ (i,j) \in E\left(G\right)$.
\end{defi}

 For a given state in this probabilistic model, there is a probability associated to each measurement 
event $\Pi_i$. 

 \begin{defi}
A behavior for the contextuality scenario $G$ is a vector
\be p \in \mathbb{R}^{\left|V\left(G\right)\right|}\ee 
such that $p_i \in [0,1]$ for every $i$ and $p_i+p_j \leq 1$ whenever $(i,j) \in E\left(G\right).$
 \end{defi}

  The set of possible behaviors depends on 
the physical theory used to describe the system. We will describe this set in detail for classical probability theory,
 quantum theory and general probability theories satisfying the E-principle.

\subsection{Classical  Realizations}

\begin{defi}
A classical realization for $G$ is given by a probability space $(\Omega,\Sigma, \mu )$, 
where $\Omega$ is a sample space, 
$\Sigma$ a $\sigma -$algebra and $\mu$ a probability measure in $\Sigma$ and 
for each $i \in V\left(G\right)$ a set $A_i \in \Sigma$
 such that $A_i \cap A_j = \varnothing$ if $(i,j) \in E\left(G\right)$. 
 For each $i$ the probability of outcome associated to $\Pi_i$ is 
\be p_i=\mu\left(A_i\right).\ee
The  behaviors that can be written in this form are called \emph{classical behaviors}. The set of all classical behaviors will be denoted by
$\mathcal{E}_C(G)$.
\end{defi}

The behaviors outside the classical set are called \emph{contextual}. 
The classical set   $\mathcal{E}_C(G)$ is a polytope, and, incidentally, this set is a well-known convex polytope in computer science literature, 
the \emph{stabilizer set} of $G$,  denoted by $\mathrm{STAB}(G)$ \cite{Knuth93,Rosenfeld67,Amaral14, AT17}. 

\begin{prop}
\label{thm:stable_clas}
The set  $\mathcal{E}_C\left(G\right)$ is equal to the stable set  $\mathrm{STAB}\left(G\right)$.
\end{prop}

Once more, since the set $\mathcal{E}_C(G)$ is a polytope, it can be characterized by a finite set
of linear inequalities which provide necessary and sufficient conditions for membership in this set.

\begin{defi}
\label{defiexclusivityinequality}
 A \emph{noncontextuality inequality} is a linear inequality 
 \be \sum \gamma_i p_i \leq b,\ee
 where all $\gamma_i$ and $b$ are real numbers, which is 
 satisfied by all elements of  the classical polytope 
 $\mathcal{E}_C(G)$ and violated by some contextual behavior.
 A noncontextuality inequality is called \emph{tight} if it is satisfied at equality for at least one classical behavior,
 and it is called  \emph{facet-defining} if it defines a  non-trivial facet of the classical polytope 
 $\mathcal{E}_C(G)$.
\end{defi}

\subsection{Quantum Realizations}

\begin{defi}
A quantum realization for $G$ is given by a density matrix $\rho$ acting  in a Hilbert space $\mathcal{H}$ and for each $i \in V\left(G\right)$ 
a projector $P_i$ acting
in $\mathcal{H}$ such that $P_i$ and $P_j$ are orthogonal  if $(i,j) \in  E\left(G\right).$ 
For each $i$ the probability of the 
outcome $i$ is 
\be p_i=\mbox{Tr}\left(P_i\rho\right).\ee
The behaviors that can be written in this form are called \emph{quantum behaviors}. The set of all quantum behaviors will be denoted by 
$\mathcal{E}_Q(G)$.
\end{defi}

This set is a well-known convex body in computer science literature, the \emph{theta body}
of $G$,  denoted by $\mathrm{TH}(G)$ \cite{Knuth93,Rosenfeld67,Amaral14}.  It is not a polytope in general.

\begin{prop}
\label{teo:q_TH}
 The set  $\mathcal{E}_Q\left(\mathrm{G}\right)$ is equal to the theta body $\mathrm{TH}\left(\mathrm{G}\right)$.
\end{prop}

If we fix a basis for $\mathcal{H}$ and consider all matrices diagonal in this basis we recover the classical models.
Hence 
\be \mathcal{E}_C(G)\subset \mathcal{E}_Q(G).\ee 
This also follows from the known fact that $\mathrm{STAB}\left(G\right) \subset \mathrm{TH}\left(G\right).$

\subsection{E-Principle Realizations}

\begin{defi}
An E-principle realization for $G$ is given by a state in a generalized  probabilistic model and for each $i \in V\left(G\right)$ a measurement event in this general theory such that the corresponding  probabilities satisfy the E-principle.
 The behaviors obtained in this way are called \emph{E-principle behaviors}.
 The set of all E-principle distributions will be denoted by   $\mathcal{E}_E(G)$.
\end{defi}

The set of E-principle behaviors 
    is also a polytope. 
This set is a well
known convex polytope in computer science literature, the \emph{clique stable set} of $G$,  denoted by $\mathrm{QSTAB}(G)$ \cite{Knuth93,Rosenfeld67}.

It is a known fact from computer science literature that $TH(G) \subset QSTAB(G)$, which is equivalent
to $\mathcal{E}_Q(G) \subset \mathcal{E}_E(G)$. This was also proven in references \cite{CSW14, FSABCLA12}.

\begin{teo}
\label{teoquantume}
 The quantum distributions satisfy the E-principle.
\end{teo}

In fact, in quantum theory, exclusive events are associated to orthogonal projectors. Hence, if $\{e_i\}$ is a set of mutually
 exclusive events, a quantum realization will provide a set $\{P_i\}$ of mutually orthogonal projectors. As a consequence
 we have 
 \be \sum_iP_i \leq I\ee
 and hence
  \be \sum_ip_i=\sum_iTr\left(P_i\rho\right) \leq Tr\left(\rho\right) \leq 1.\ee

\begin{prop}
\label{teo:e_qstab}
 The E-principle set $\mathcal{E}_\mathrm{E}\left(\mathrm{G}\right)$ is equal to $\mathrm{QSTAB}\left(\mathrm{G}\right)$.
\end{prop}

%\begin{dem}
 %A subset of vertices $Q \subset \mathrm{V}\left(\mathrm{G}\right)$ is a clique in the exclusivity graph $\mathrm{G}$  if, 
 %and only if, the corresponding set of measurement 
 %events is pairwise exclusive. Hence, the conditions
 %$p_i \geq 0$ and $\sum_{i \in Q} p_i \leq 1$ are exactly equivalent to the constraints that each $p_i$ represents a probability and that the 
 %E-principle is satisfied.
%\end{dem}

For a proof of Props. \ref{thm:stable_clas}, \ref{teo:q_TH} and \ref{teo:e_qstab}, see Refs. \cite{CSW14, Amaral14, AT17}.

\subsection{Contextuality Quantifyers in the Exclusivity-Graph Approach}
\label{sectionexclusivityquantifyers}
We look now for functions $X: \mathcal{E}_E\left(G\right) \rightarrow \mathbb{R}_+$ that give a 
quantitative characterization of contextuality in the exclusivity graph approach. 
We still lack a proper parametrization of a physically relevant set of free operations in this case, but such a set must necessarily 
contain the \emph{relabelling operations}.

\begin{defi}
 A \emph{relabeling operation} $\mathcal{T}_{\phi}$ in the scenario $G$ is defined by
 \be
 \mathcal{T}_{\phi}\left(p\right)_i=p_{\phi(i)},
 \ee
where  $\phi$ is a graph isomorphism of $G$. 
\end{defi}

Notice that this operation corresponds to the permutations of the entries of $p$ consistent with the 
exclusivity relations given by $G$.

We demand that any contextuality monotone $X$ be preserved under the action of relabeling operations: 
\be
X\left(\mathcal{T}_{\phi}\left(p\right)\right) = X\left(p\right).
\ee
Moreover, some additional properties are also desirable:

\begin{enumerate}

\item  \emph{Faithfulness:} For all $p \in \mathcal{E}_C(G)$, $X\left(p\right)=0$.

 %The set of allowed transformations is presented in reference \cite{}.
%General transformations can be written in the block form 
%$$\mathcal{T}= \left[\begin{array}{ccc}
 %                \alpha_{11} T_{11} & \cdots& \alpha_{1N} T_{1N}\\
 %                \vdots&\ddots& \vdots\\
  %               \alpha_{N1}T_{N1}&\ldots &\alpha_{NN}T_{NN}          
   %            \end{array}\right]$$
%where $T_{ij}$ is a stochastic matrix taking a distribution on context $j$ to a distribution in context $i$ and
%$\sum_j \alpha_{ij}=1$. 

%Since a stochastic transformation decreases the norm of a vector, monotonicity is valid for $D$.

\item  \emph{Additivity:} 
First we consider a scenario  such that its  exclusivity graph $G$ consists of two
connected components $G_1$ 
and
$G_2$. The behaviors for $G$ are formed by the list of probabilities $p_1$ for the scenario given by $G_1$ 
followed by the list of  probabilities $p_2$
for the scenario given by $G_2$. It follows that   any behavior $p$  in $G$ is the juxtaposition of a 
behavior $p_1$  for $G_1$ and a behavior $p_2$ for $G_2$. Such behaviors will be denoted by $p_1\&p_2$.
The quantifier $X$ should be such that 
\be
X\left(p_1\&p_2\right)\leq X\left(p_1\right)+X\left(p_2\right).
\ee
One may also require that equality holds.

Another kind of operation we can apply to two scenarios is considering the set of  events where an event  in $G_1$ and an event in
$G_2$ are true,  with the restriction that they should be independent. 
This implies that a behavior for $G$
is the tensor product of one behavior for $G_1$ with a behavior for $G_2$. For this kind of operation,
subadditivity of $X$ should also hold. 
%\begin{eqnarray*}
%X\left(P \otimes P_{\mathcal{C}'}\right)\leq D\left(P \otimes P_{\mathcal{C}'}, Q_{\mathcal{C}} \otimes Q_{\mathcal{C}'}\right) &=& \sum_{k,l} |p_kp'_l - q_kq'_l|\\
%&=&
%\sum_{k,l} |p_kp'_l - p_kq'_l +p_kq'_l- q_kq'_l|\\
%&\leq &\sum_{k,l} p_k|p'_l - q'_l| + q'_l|p_k - q_k|\\
%&=&D\left(P , Q_{\mathcal{C}} \right)+D\left( P_{\mathcal{C}'}, Q_{\mathcal{C}'}\right)\\
%&=&X\left(P\right)+X\left(P_{\mathcal{C'}}\right).
%\end{eqnarray*}
\be
X\left(p_1 \otimes p_2\right)\leq X\left(p_1\right)+X\left(p_2\right).
\ee

\item  \emph{Convexity:} If a  behavior can be written as $p=\sum_i \pi_ip^i$, where  $\pi_i \in [0,1]$ and each $p^i$ is a behavior  for the same scenario, then 
\be X(p) \leq \sum_i \pi_i X\left(p^i\right).\ee

\item  \emph{Continuity:} $X\left(p\right)$ is a continuous function of $p$.

\end{enumerate}

In this approach, we can also define  contextuality quantifiers based on the geometry of the set o distributions. 

\begin{defi}Given a distance $D$
 in $\mathbb{R}^{\left|V\left(G\right)\right|}$, 
we define the \emph{contextuality distance}
\be \mathcal{D}(p)=\frac{1}{\left|V\left(G\right)\right|}\min _{q \in \mathrm{STAB}(G)} D\left(p , q \right). \label{eqdefdist_e}\ee
\end{defi}

\begin{defi}
The \emph{robustness} of a behavior $p$ is defined as
\be
\label{eq:rob_exc}
\mathcal{R}\left(p\right)= \min \left\{\lambda \left|  \left(1-\lambda\right) p + \lambda p^{NC} \in \mathrm{STAB}\left(G\right)\right.\right\},
\ee
where $B^{NC}$ is an arbitrary noncontextual behavior.
\end{defi}

\begin{defi}
The \emph{contextual fraction} of a behavior $B$ is defined as
\be
\label{eq:cont_frac_exc}
\mathcal{F}\left(B\right)= \min \left\{\lambda \left|B=  \lambda B' + \left(1-\lambda\right)B^{NC}\right.\right\},
\ee
where $B^{NC}$ is an arbitrary noncontextual behavior.
\end{defi}

%\be X_2(p)=\frac{1}{N}\min _{P^{NC}} \sum_{J\in \mathcal{C}} D_{1}\left(P_J  \middle\| P^{NC}_I \right), \ q \in \mathcal{NC}(\Gamma), \label{eqdefdist2}\ee

%\be X_3(p)=\sup_{P(J)} \min _{P^{NC}} \sum _{J\in \mathcal{C}}\ p(J)\ D_{1}\left(P_J  \middle\| P^{NC}_I \right), \ q \in \mathcal{NC}(\Gamma), \label{eqdefdist3}\ee
%\begin{equation}\label{ContDist}

\begin{teo}
\begin{enumerate}
 \item $\mathcal{D}, \mathcal{F}$  and $\mathcal{R}$ are faithful, convex, subadditive under products and continuous;
  \item $\mathcal{D}, \mathcal{F}$  and $\mathcal{R}$ are preserved under relabeling operations;
 \item $\mathcal{F}$ and $\mathcal{R}$ are monotonous under all linear operations that preserve $\mathrm{STAB}\left(G\right);$
 \item $\mathcal{F}$ and $\mathcal{R}$ can be computed via linear programming.
\end{enumerate}

\end{teo}

The proofs are analogous to the ones presented for the compatibility-hypergraph approach, and hence we do not repeat them here.

  \subsection{Connection to graph invariants}

%To obtain necessary and sufficient conditions for membership in $\mathcal{E}_C(G)$, we have to find all 
%tight noncontextuality inequalities for $G$. This is a difficult problem, in general.

Given a  graph $G$, consider the sum of probabilities
\begin{equation}
\beta  = \sum_{i \in V} \gamma_i p_i.
\end{equation}
We can use this sum and graph invariants to provide  necessary conditions to membership in $\mathcal{E}_C(G)$,
$\mathcal{E}_Q(G)$ and $\mathcal{E}_E(G)$. 
%To derive these
%conditions we need to identify what are the maximum values of $\beta$ for each of classical, quantum and E-principle 
%realizations, which will be denoted respectively by
Let $\beta_C$, $\beta_Q$ and $\beta_E$ be the maximum values of $\beta$ for each of classical, quantum and E-principle 
realizations, respectively.
%Naturally, by theorem \ref{teoquantume} and the fact that $\mathcal{E}_C(G)\subset \mathcal{E}_Q(G)$, we have
%\be \beta_C \leq \beta_Q \leq \beta_E.\ee

%The inequality
%\begin{equation}
% \sum_{i \in V} \gamma_i p_i \leq \beta_C
%\end{equation}
%is  a \emph{noncontextuality inequality} as long as $\beta_C < \beta_E$ and
%\begin{equation}
 %\sum_{i \in V} \gamma_i p_i \leq \beta_Q
%\end{equation}
%is a necessary condition for membership in $\mathcal{E}_Q(G)$.

%Also in the exclusivity-graph approach, graph invariants can be used to
%calculate $\beta_C$ and $\beta_Q$ and $\beta_E$.

\begin{prop}[Cabello, Severini, and Winter, 2010]
Given a graph $G$,
\be \beta_C=\alpha(G,\gamma), \ \beta_Q= \vartheta(G,\gamma),  \ \beta_E=\alpha^*(G, \gamma)\ee
where $\alpha(G, \gamma)$ is the weighted independence number of $G$, $\vartheta(G, \gamma)$ is  the weighted  Lov\'asz number of $G$ and 
$\alpha^*(G, \gamma)$ is the weighted  fractional-packing number of $G$. 
\label{teoqboundexclusivity}
\end{prop}

This result follows directly from the observation that $\mathcal{E}_C(G)=\mathrm{STAB}(G), \ \mathcal{E}_Q(G)=TH(G)$ and 
$\mathcal{E}_E(G)=QSTAB(G)$ and the well known fact from computer science literature that 
$\alpha(G, \gamma), \ \vartheta(G, \gamma),  \ \alpha^*(G,\gamma)$ are the maximum values of $\sum_i \gamma_ip_i$
over $STAB(G), \ TH(G),$ and $ \ QSTAB(G)$ respectively \cite{Rosenfeld67, Knuth93, CSW14}.

In some situation we can use the connection with graph theory to calculate the distance defined in Eq.  \ref{eqdefdist}.
This is the case for the $n$-cycle inequalities.

\subsection{A new version of the $n$-cycle inequalities}
\label{exnewncycle}
 The simplest exclusivity  graph for which $\beta_C < \beta_Q$ is the pentagon \cite{CDLP13}.
 It can be proven by inspection that $\beta_C=2$. In this case, there is only one non-trivial facet-inducing inequality for 
the stable set, given by 
\be \sum_ip_i=2.\ee
 
 The quantum bound  is given by the Lov\'asz number $\vartheta(C_5)=\sqrt{5}$, as shown by 
 Lov\'asz original calculation \cite{Lovasz79}.
  The maximum value obtained  with E-distributions  is $\frac{5}{2}$, which can be reached when all events 
  have probability equal
  to $\frac{1}{2}$.
 
When $G$ is any $n$-cycle with $n$  odd, we can also prove by inspection that the classical bound is $\beta_C=\frac{n-1}{2}$.
The quantum bound can also be explicitly calculated, and we have that 
 \be \beta_Q=\frac{n\cos\left(\frac{\pi}{n}\right)}{1+\cos\left(\frac{\pi}{n}\right)}, \ee which is equal to $\sqrt{5}$ for $n=5$. 
 The maximum obtained with E-distributions   is $\frac{n}{2}$, which can be reached when all events 
  have probability equal
  to $\frac{1}{2}$.
 
% If $n$ is even, $C_n$ is a bipartite graph, and the vertices in one bipartition define a maximal  independent set.
% The parts have the same size, and hence
 %the classical bound is $\frac{n}{2}$. The distribution that assigns probability $\frac{1}{2}$ to all 
 %vertices realizes the bound $\beta_E$, which is then equal to $\beta_C$.
 %The quantum bound $\beta_Q$ is sandwiched between $\beta_C$ and $\beta_E$ and hence we conclude that
 %$\beta_Q$ is also equal to $\frac{n}{2}$.

For any odd $n$, there is only one non-trivial facet-inducing inequality  for $STAB\left(C_n\right)$, given by
\be \sum_ip_i=\frac{n-1}{2}.\ee
Hence, each contextual distribution violates only one facet-defining inequality, and  the distance
of such a point to the set of non-contextual distributions is equal to the distance of this distribution
to the hyperplane defining the facet. 
Then, we have

\begin{teo} The distance with respect to the $\ell_r$ norm from  $p$  to $STAB\left(C_n\right)$ is
\be D_r(x)= \frac{\sum_ip_i -\frac{n-1}{2}}{\sqrt[q]{n}} \ee
whith
\be \frac{1}{r}+\frac{1}{q}=1.\ee
In particular, for the   $\ell_2$ norm we have
\be D(p)= \frac{\sum_ip_i -\frac{n-1}{2}}{\sqrt{n}},\ee
for  the sum norm we have \be D_1(p)= \sum_ip_i -\frac{n-1}{2},\ee and for  the maximum norm we have
\be D_{\infty}(p)= \frac{\sum_ip_i -\frac{n-1}{2}}{n}.\ee
\end{teo}

Once more, the maximum distance for quantum and E-principle behaviors can be computed from graph invariants of $C_n$.

 \begin{teo}
 The maximum value of $\mathcal{D}_r$ for the $n$-cicle  scenario, $n$ odd, attainable with quantum behaviors is
 \be  \frac{\vartheta\left( C_n\right)-\alpha\left( C_n\right)}{n\sqrt[r]{n}}.\ee
  The maximum value of $\mathcal{D}_r$ for the $n$-cicle scenario attainable with E-principle behaviors is
 \be  \frac{\alpha^*\left( C_n\right)-\alpha\left( C_n\right)}{n\sqrt[r]{n}}.\ee

 \end{teo}

The results above  show that when the point violates only one contextualtiy  inequality, the distance is related with that 
inequality. In particular, the maximum distance is given when the violation is maximum.
The same approach is valid for any point that violates only one inequality. Unfortunately,  is not always the case, since the  $STAB(G)$ polytope
has an intricate structure for complicated graphs. See Appendix \ref{ap:2ineq} for a example of a scenario with this property.

\section{Conclusion}
\label{sectionconclusion}
The complete description of a contextuality scenario is a difficult problem in general, since the complexity of the set 
of distributions grows enormously with the number of measurements  avaible. Nonetheless, we can use several geometric features of
this set to help us in this task. The graph approach to  contextuality is an essential tool, since we can translate several problems in contextuality
to problems already studied in graph theory. In particular, we can then identify several well known convex sets  that appear
in contextuality with well known convex sets from graph theory literature. 

The identification of these sets gives us a beautiful geometry which  can be explored. We can, for example, use it to define 
contextuality quantifiers based on the geometrical distances in such convex sets. This definition is important in
the resource theory of contextuality.
The advantages of our definition is threefold: we can connect our quantifiers with graph invariants;
they can be computed more efficiently then the quantifiers based in 
relative entropy; they  can also be applied to 
the exclusivity graph approach to contextuality, where previous quantifiers do not fit.
%Understand and use the potential application of quantum theory to process information, including quantum communication
%and quantum computing, is essential for the development of technology in many areas and quantum contextuality may play a important role.

\begin{acknowledgments}
We thank S. Abramsky, L. Aolita,  R. S. Barbosa, S. G. A. Brito, A. Cabello, P. Horodecki, M. Kleinmann, J.-\AA{}. Larsson, S. Mansfield and  M. T. Quintino,
for valuable discussions. We acknowledge financial support from the Brazilian ministries MEC and MCTIC, and  agencies CAPES, CNPq, FAPEMIG, FAEPEX, INCT. 
\end{acknowledgments}

\include{ap}

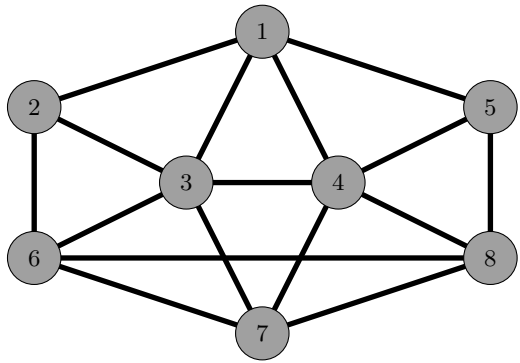
\begin{figure}[h!]
\begin{center}
\definecolor{aqaqaq}{rgb}{0.6274509803921569,0.6274509803921569,0.6274509803921569}
\begin{tikzpicture}[line cap=round,line join=round,>=triangle 45,x=1cm,y=1cm]
%\clip(-10.697697839381949,-8.091819550188015) rectangle (5.417368448427076,4.823091730112817);
\draw [line width=2pt] (-7,2)-- (-4,3);\draw [line width=2pt] (-4,3)-- (-1,2);
\draw [line width=2pt] (-1,2)-- (-1,0);\draw [line width=2pt] (-1,0)-- (-4,-1);
\draw [line width=2pt] (-4,-1)-- (-7,0);\draw [line width=2pt] (-7,0)-- (-7,2);
\draw [line width=2pt] (-5,1)-- (-3,1);\draw [line width=2pt] (-3,1)-- (-4,-1);
\draw [line width=2pt] (-5,1)-- (-4,-1);\draw [line width=2pt] (-1,0)-- (-3,1);
\draw [line width=2pt] (-7,0)-- (-5,1);\draw [line width=2pt] (-5,1)-- (-7,2);
\draw [line width=2pt] (-5,1)-- (-4,3);\draw [line width=2pt] (-3,1)-- (-4,3);
\draw [line width=2pt] (-3,1)-- (-1,2);\draw [line width=2pt] (-7,0)-- (-1,0);
\draw [fill=aqaqaq] (-7,2) circle (10pt);
\draw[color=black] (-7,2) node {$2$};
\draw [fill=aqaqaq] (-7,0) circle (10pt);
\draw[color=black] (-7,0) node {$6$};
\draw [fill=aqaqaq] (-5,1) circle (10pt);
\draw[color=black] (-5,1) node {$3$};
\draw [fill=aqaqaq] (-3,1) circle (10pt);
\draw[color=black] (-3,1) node {$4$};
\draw [fill=aqaqaq] (-1,2) circle (10pt);
\draw[color=black] (-1,2) node {$5$};
\draw [fill=aqaqaq] (-1,0) circle (10pt);
\draw[color=black] (-1,0) node {$8$};
\draw [fill=aqaqaq] (-4,-1) circle (10pt);
\draw[color=black] (-4,-1) node {$7$};
\draw [fill=aqaqaq] (-4,3) circle (10pt);
\draw[color=black] (-4,3) node {$1$};
\end{tikzpicture}
\end{center}
\caption{A graph $G$ with a behavior violting two facet defining inequalities of the stable set polytope. }
\label{fig1}
\end{figure} 


%
% BibTeX users please use
% \bibliographystyle{}
% \bibliography{}
%\begin{thebibliography}{}
%Author, \textit{Book title} (Publisher, place year) page numbers
% etc
%\end{thebibliography}

%\renewcommand{\bibname}%
%{\textsf{Bibliografia}}
\bibliographystyle{abbrv}
\bibliography{biblio}

\end{document}

%% file: comandos1.tex
\newcommand{\De}[1]{\left[ #1 \right]}

\newcommand{\ket}[1]{\left| #1 \right\rangle}
\newcommand{\bra}[1]{\left\langle #1 \right|}
\newcommand{\braket}[2]{\left\langle #1 \mid #2 \right\rangle}

\newcommand{\tr}{\mathrm{Tr}}

\newcommand{\textLP}{\texttt}

\newtheorem{teo}{Theorem}

\newtheorem{cor}{Corollary}
\newtheorem{prop}{Proposition}

\newtheorem{defi}{Definition}
\newtheorem{prin}{Principle}

\newenvironment{dem}{\vspace{1ex}\noindent{\textit{Proof.}}\hspace{0.3em}}{\hfill\qed\vspace{1ex}}

%% file: ap.tex
\section{Appendix}

\subsection{Proof of Theorem \ref{teo:prop_entropic}}
\label{app:entropic}

\setcounter{teo}{0}
\begin{teo}
\begin{enumerate}
\item $E_{max}$ is a contextuality monotone for the resource theory of contextuality defined by the noncontextual wirings;
\item $E_u$ is a contextuality monotone for the resource theory of contextuality defined by post-processing operations and a subclass of pre-processing operations;
\item $E_{max}$ and $E_u$ are faithful, additive, convex, continuous, and preserved under relabellings of inputs and outputs.
\end{enumerate}
\end{teo}

\noindent \textit{Proof.} \hspace{0.3em} It was shown in Refs. \cite{GHHHHJKW14} that  $E_{max}$ and $E_u$ are faithful, additive, convex, continuous, and preserved under
relabellings of inputs and outputs. The proof that $E_max$ is monotonous under noncontextual wirings was given in 
Ref. \cite{ACTA17}. We now prove that $E_u$ is monotonous under post-processing  operations and a restricted class of pre-processing operations.

\subsubsection{Monotonicity of $E_u$ under post-processing operations}
A \emph{post-processing} operation $\mathcal{O}$ takes the behavior $B=\left\{p_C(s)\right\}$ to the behavior 
$\mathcal{O}(B)=\left\{p^f_C\left(s'\right)\right\}$ with
\be
p^f_C(s') = \sum_s O^{NC}_s\left(s'\right)p_C(s)
\label{eq:post}
\ee
where $O^{NC}_s$ is a noncontextual behavior with input $s$ and output $s'$. It was shown in Ref. \cite{ACTA17} that 
post-processing operations preserve the set of noncontextual behaviors.

Let $B^* = \left\{q_C(s)\right\}$ be the behavior achieving the minimum for $B$ in Eq. \eqref{HorDist3}, that is,
\be 
E_{u}\left(B \right) = \frac{1}{N} \ \sum _{C\in \mathcal{C}}\ D_{\mathrm{KL}}\left(p_C  \middle\| q_C \right),
\ee 
and $\mathcal{O}\left(B^*\right) = \left\{q^f_C\left(s'\right)\right\}$ its image under a  post-processing operation $\mathcal{O}$. Then, we have

\begin{eqnarray}
 E_u\left(\mathcal{O}(B)\right) & \leq & \frac{1}{N}\sum_{C,s'} p^f_C\left(s'\right) \log\left(\frac{p^f_C\left(s'\right)}{q^f_C\left(s'\right)}\right) \label{eq:entropic_1}\\
 &=& \frac{1}{N}\sum_{C,s'} \left[\sum_s O^{NC}_s\left(s'\right)p_C(s)\right] \log\left(\frac{\sum_s O^{NC}_s\left(s'\right)p_C(s)}{\sum_s O^{NC}_s\left(s'\right)q_C(s)}\right) \label{eq:entropic_2}\\
&\leq & \frac{1}{N}\sum_{C,s', s} O^{NC}_s\left(s'\right)p_C(s) \log\left(\frac{p_C(s)}{q_C(s)}\right) \label{eq:entropic_3}\\
&\leq & \frac{1}{N}\sum_{C, s}p_C(s) \log\left(\frac{p_C(s)}{q_C(s)}\right) \label{eq:entropic_4}\\
 &=& E_u(B) \label{eq:entrop_5}.
\end{eqnarray}
Eq. \eqref{eq:entropic_1} follows from the fact that $\mathcal{O}\left(B^*\right)$ is a noncontextual behavior, 
Eq. \eqref{eq:entropic_2} follows from Eq. \eqref{eq:post}, Eq. \eqref{eq:entropic_3} follows from the log-sum inequality,
Eq. \eqref{eq:entropic_4} follows from the fact that $\sum_r O^{NC}_s\left(s'\right)=1$ and Eq. \eqref{eq:entrop_5}
follows from the fact that $B^*$ is the behavior achieving the minimum  in Eq. \eqref{HorDist3}.

\subsubsection{Monotonicity of $E_u$ under restricted pre-processing operations}
A \emph{pre-processing} operation $\mathcal{I}$ takes the behavior $B=\left\{p_C(s)\right\}$ to the behavior 
$\mathcal{I}(B)=\left\{p^f_{C'}(s)\right\}$ with
\be
p^f_{C'}(s) = \sum_C p_C(s)I^{NC}_{C'}(C)
\label{eq:pre}
\ee
where $I^{NC}_C$ is a noncontextual behavior with input $C'$ and output $C$. It was shown in Ref. \cite{ACTA17} that 
pre-processing operations preserve the set of noncontextual behaviors.

Let $\mathcal{I}$ be a pre-processing operation such that the number of contexts is preserved and such that
\be
\sum_{C'} I^{NC}_{C'}(C) \leq 1 \ \forall \ C.
\label{eq:pre_esp}
\ee
Let $B^* = \left\{q_C(s)\right\}$ be the behavior achieving the minimum for $B$ in Eq. \eqref{HorDist3}  
and $\mathcal{I}\left(B^*\right) = \left\{q^f_{C'}(s)\right\}$ its image under a pre-processing operation $\mathcal{I}$. Then, we have

\begin{eqnarray}
 E_u\left(\mathcal{I}(B)\right) & \leq & \frac{1}{N}\sum_{C',s} p^f_{C'}(s) \log\left(\frac{p^f_{C'}(s)}{q^f_{C'}(s)}\right) \label{eq:entropic_1_pre}\\
 &=& \frac{1}{N}\sum_{C',s} \left[\sum_C p_C(s)I^{NC}_{C'}(C)\right] \log\left(\frac{\sum_C p_C(s)I^{NC}_{C'}(C)}{\sum_C q_C(s)I^{NC}_{C'}(C)}\right) \label{eq:entropic_2_pre}\\
&\leq & \frac{1}{N}\sum_{C',C, s}  p_C(s)I^{NC}_{C'}(C) \log\left(\frac{p_C(s)}{q_C(s)}\right) \label{eq:entropic_3_pre}\\
&\leq & \frac{1}{N}\sum_{C, s}p_C(s) \log\left(\frac{p_C(s)}{q_C(s)}\right) \label{eq:entropic_4_pre}\\
 &=& E_u(B) \label{eq:entrop_5_pre}.
\end{eqnarray}
Eq. \eqref{eq:entropic_1_pre} follows from the fact that $\mathcal{I}\left(B^*\right)$ is a noncontextual behavior, 
Eq. \eqref{eq:entropic_2_pre} follows from Eq. \eqref{eq:post}, Eq. \eqref{eq:entropic_3_pre} follows from the log-sum inequality,
Eq. \eqref{eq:entropic_4_pre} follows from Eq. \eqref{eq:pre_esp} and Eq. \eqref{eq:entrop_5}
follows from the fact that $B^*$ is the behavior achieving the minimum  in Eq.  \eqref{HorDist3}.

 \hfill\qed\vspace{1ex}

\subsection{Proof of Theorem 2}
\label{ap:dist}

\begin{teo}
\begin{enumerate}
\item $\mathcal{D}_{max}$ is a contextuality monotone for the resource theory of contextuality defined by the noncontextual wiring operations;
\item $\mathcal{D}_u$ is a contextuality monotone for the resource theory of contextuality defined by post-processing operations and a subclass of pre-processing operations;
\item $\mathcal{D}$, $\mathcal{D}_{u}$ and $\mathcal{D}_{max}$ are faithful, additive, convex, continuous, and preserved under relabellings of inputs and outputs.
\end{enumerate}
\end{teo}

\noindent \textit{Proof.} The several steps of the proof are presented in the subsections below.

\subsubsection{Monotonicity under Free Operations of Contextuality}

Let $\mathcal{W}$ be a noncontextual wiring, as defined in Ref. \cite{ACTA17}. Such an operation
takes the behavior $B=\left\{p_C(s)\right\}$ to the behavior $\mathcal{W}\left(B\right)=\left\{p^{f}_{C'}\left(s'\right)\right\}$ defined as
\be
p^f_{C'}\left(s'\right) = \sum_{C, s} O^{C', C}_s\left(s'\right) p_C(s)I_{C'}(C)
\ee
where $\left\{I_{C'}(C)\right\}$ is a pre-processing noncontextual behavior with inputs $C'$ and outputs $C$ and 
$\left\{O^{C', C}_s\left(s'\right)\right\}$
is a post-processing noncontextual behavior  that may also depend on the pre-processing, but in a restricted way in order to 
preserve the set of noncontextual behaviors (see Ref. \cite{ACTA17} for details). Notice that post and pre-processing operations are particular cases of
noncontextual wirings.

We first prove that $D_{max}$ is monotonous under pre-processing operations. Given a behavior $B$, let $B^*=\left\{q_C(s)\right\}$ be the behavior achieving the minimum in equation \eqref{eqdefdist3}, that is,
\be \mathcal{D}_{max}\left(B\right)=\max_{\pi} \sum _{C\in \mathcal{C}}\ \pi(C)\ D\left(p_C, q_C \right),\ee
and $\mathcal{I}\left(B^*\right) = \left\{q^f_{C'}\left(s'\right)\right\}$ its image under  $\mathcal{I}$.
Then,
\begin{eqnarray}
 \mathcal{D}_{max}\left(\mathcal{I}(B)\right) & \leq & \max_{C'} \sqrt[p ]{\sum_{s}\left(p^ f_{C'}\left(s\right)-q^f_{C'}\left(s\right)\right)^p}\label{eq:dmax_1}\\
 & = & \max_{C'} \sqrt[p]{\sum_{s}\left[\sum_{ C}I_{C'}(C) \left(p_{C}\left(s\right)-q_{C}\left(s\right)\right)\right]^ p}\label{eq:dmax_2}\\
 &\leq & \max_{C'} \sum_{ C}  I_{C'}(C)  \sqrt[p]{ \sum_{s}\left(p_{C}\left(s\right)-q_{C}\left(s\right)\right)^p}\label{eq:dmax_3}\\
 & \leq &  \max_{C}  \sqrt[p]{ \sum_{s}\left(p_{C}\left(s\right)-q_{C}\left(s\right)\right)^p}\label{eq:dmax_4}\\
 &=&\mathcal{D}_{max}\left(B\right). \label{eq:dmax_5}
 \end{eqnarray}
 
Eq. \eqref{eq:dmax_1} follows from the fact that $\mathcal{I}\left(B^*\right)$ is a noncontextual behavior, 
Eq. \eqref{eq:dmax_2} follows from Eq. \eqref{eq:pre}, Eq. \eqref{eq:dmax_3} follows from Minkowski inequality,
Eq. \eqref{eq:dmax_4} follows from the fact that the mean is less or equal than the maximum, and Eq. \eqref{eq:dmax_5}
follows from the fact that $B^*$ is the behavior achieving the minimum  in Eq.  \eqref{HorDist3}.

We now prove that $D_{max}$ is monotonous under post-processing operations, we notice that if $\mathcal{O}$ is a post-processing operation, then for each context $C$ there is a 
stochastic matrix 
$M^C$ such that
\be p^f_C=M^Cp_C. \label{eq:post_matrix}\ee
Let $B^*=\left\{q_C(s)\right\}$ be the behavior achieving the minimum in equation \eqref{eqdefdist3},
and $\mathcal{O}\left(B^*\right) = \left\{q^f_{C'}\left(s'\right)\right\}$ its image under  $\mathcal{O}$.

\begin{eqnarray}
 \mathcal{D}_{max}\left(\mathcal{O}(B)\right) & \leq & \max_{C} \left\|p^ f_{C}-q^f_{C}\right\|_{\ell_p}\label{eq:dmax_post_1}\\
 & = & \max_{C} \left\| M^C \left(p_C-q_C\right)\right\|_{\ell_p}\label{eq:dmax_post_2}\\
 &\leq & \max_{C} \left\|  \left(p_C-q_C\right)\right\|_{\ell_p}\label{eq:dmax_post_3}\\
 &=&\mathcal{D}_{max}\left(B\right). \label{eq:dmax_post_4}\end{eqnarray} 
Eq. \eqref{eq:dmax_post_1} follows from the fact that $\mathcal{O}\left(B^*\right)$ is a noncontextual behavior, 
Eq. \eqref{eq:dmax_post_2} follows from Eq. \eqref{eq:post_matrix}, Eq. \eqref{eq:dmax_3} follows from the fact that a stochastic matrix must satisfy
\be \left\|Mx\right\|_{\ell_p} \leq \|x\|_{\ell_p} ,\ee
and Eq. \eqref{eq:dmax_4}  from the fact that $B^*$ is the behavior achieving the minimum  in Eq.  \eqref{HorDist3}.
 
 The quantifiers $\mathcal{D}$ and $\mathcal{D}_u$ are not monotonous under the entire class of noncontextual wirings. Nevertheless
 they are monotonous under output operations and under the restricted class of input operations defined in Eq. \eqref{eq:pre_esp}. The proofs
 are analogous to the ones presented for $\mathcal{E}_u$ and can also be found in Ref. \cite{BAC17}, for the restricted case of Bell scenarios.

Reversible wirings correspond to permutations of inputs and outputs and correspond to a permutation of the entries of $B$.
Hence, these transformations preserve $\mathcal{D}$, $\mathcal{D}_u$ and $\mathcal{D}_{max}$. 

\subsubsection{Additivity under juxtaposition}
% Any behavior  in the juxtaposition can be written in the form $B=B_1\& B_2$, where 
% $B_1$ is a behavior for $H_1$ and $B_2$ is a behavior for $H_2$. 
Let $B=B_1 \& B_2$ be the juxtaposition of $B_1$ and $B_2$, and $B^*_1$ and $B^*_2$ the non-contextual behaviors
that achieve the minimum in Eq. \eqref{eqdefdist} for $B_1$ and $B_2$ respectively. Then
\begin{eqnarray}
\mathcal{D}\left(B_1 \& B_2\right) & \leq & \left\|P_{B_1\& B_2} -P_{ B_1^*\&B_2^*}\right\|_{\ell_p}  \\
 & = & \left\|P_{B_1\&0}+P_{0\&B_2} -P_{B_1^*\&0}-P_{0\&B_2^*}\right\|_{\ell_p}\\
 &\leq & \left\|P_{B_1\&0}-P_{B_1^*\&0}\right\|_{\ell_p} + \left\|P_{0\&B_2} -P_{0\&B_2^*}\right\|_{\ell_p}\\
 &=& \left\|P_{B_1}-P_{B_1^*}\right\|_{\ell_p} + \left\|P_{B_2} -P_{B_2^*}\right\|_{\ell_p}\\
 &=&\mathcal{D}\left(B_1\right) + \mathcal{D}\left(B_2\right).
\end{eqnarray}
Equality holds for the $\ell_1$ norm. For the $\ell_\infty$ norm a similar argument shows that 
\begin{equation}
 \mathcal{D}\left(B_1 \& B_2\right) \leq \max_i \mathcal{D}\left(B_i\right).
\end{equation}
For $\mathcal{D}_u$, a similar argument proves that 
\begin{equation}
 \mathcal{D}_u\left(B_1 \& B_2\right) \leq \frac{1}{N_2} \mathcal{D}\left(B_1\right) + \frac{1}{N_1} \mathcal{D}\left(B_2\right),
\end{equation}
and for $\mathcal{D}_{max}$
\begin{equation}
 \mathcal{D}_{max}\left(B_1 \& B_2\right) \leq \max_i \mathcal{D}_{max}\left(B_i\right).
\end{equation}
\subsubsection{Sub-additivity for the tensor product} 
Although  $\mathcal{D}$ is not sub-additive under tensor products for a general distance $D$, $\mathcal{D}_u$ and $\mathcal{D}_{max}$ are additive when $D$ is defined by a $\ell_p$ norm. Let $B_1 \otimes B_2$ be the tensor product of $B_1$ and $B_2$, and $B^*_1$ and $B^*_2$ the non-contextual behaviors
that achieve the minimum in Eq. \eqref{eqdefdist2} for $B_1$ and $B_2$ respectively. Then

\begin{eqnarray}
\mathcal{D}_u\left(B_1 \otimes B_2\right) &\leq &\frac{1}{N_1 N_2} \left\| P_{B_1} \otimes P_{B_2} - P_{B_1^ *} \otimes P_{B_2^ *}\right\| \label{eq:du_tp_1}\\
%&= & \frac{1}{N_1 N_2} \left\| P_{B_1} \otimes P_{B_2} -P_{B_1^ *} \otimes P_{B_2} + P_{B_1^ *} \otimes P_{B_2} - P_{B_1^ *} \otimes P_{B_2^ *}\right\| \label{eq:du_tp_2}\\
&\leq & \frac{1}{N_1 N_2} \left\| P_{B_1} \otimes P_{B_2} -P_{B_1^ *} \otimes P_{B_2} \right\| \nonumber\\
& & \ \ \ \ \ \ \  + \frac{1}{N_1 N_2}\left\| P_{B_1^ *} \otimes P_{B_2} - P_{B_1^ *} \otimes P_{B_2^ *}\right\| \label{eq:du_tp_3}\\
&=&\frac{1}{N_1 N_2} \left\| P_{B_1}  -P_{B_1^ *}\right\|\left\|P_{B_2} \right\| \nonumber\\
&& \ \ \ \ \ \ \ + \frac{1}{N_1 N_2}\left\| P_{B_1^ *}\right\|\left\|s P_{B_2} -  P_{B_2^ *}\right\| \label{eq:du_tp_4}\\
&=&\frac{1}{N_1} \left\| P_{B_1}  -P_{B_1^ *}\right\| + \frac{1}{N_2}\left\| P_{B_2} -  P_{B_2^ *}\right\| \label{eq:du_tp_5}\\
&=& \mathcal{D}_u\left(B_1\right) + \mathcal{D}_u\left(B_2\right)\label{eq:du_tp_6}
\end{eqnarray}
Eq. \ref{eq:du_tp_1} follows from the fact that $B_1^ * \otimes B_2^ *$ is a noncontextual behavior, Eq. \ref{eq:du_tp_3} follows from the triangular inequality, 
Eq. \ref{eq:du_tp_4} follows from the multiplicativity of $\ell_p$ norms under tensor products, Eq. \ref{eq:du_tp_5} follows from the fact that $\left\|P_{B_1} \right\|=N_1$ and $\left\|P_{B_2} \right\|=N_2$, and Eq. \ref{eq:du_tp_6} follows from the fact that $B^*_1$ and $B^*_2$  are the non-contextual behaviors
that achieve the minimum in Eq. \eqref{eqdefdist2} for $B_1$ and $B_2$ respectively.

A similar argument shows that $\mathcal{D}_{max}$ is sub-additive under tensor products.

\subsubsection{Convexity}
If a behavior can be written as $B=\sum_i \pi(i)B_i$ then 
\be \mathcal{D}(B) \leq \sum_i \pi(i) \mathcal{D}\left(B_i\right).\ee
In fact,  let $B_i$ be the non-contextual behavior
achieving the minimum for $B_i^*$ in Eq. \eqref{eqdefdist}. Then
\begin{eqnarray}
  \mathcal{D}\left(\sum_i \pi(i) B_i\right) &\leq&  D\left(\sum_i \pi(i) P_{B_i}, \sum_i \pi(i) P_{B^*_i}\right) \label{eq:convexity_1}\\
  &= &\left\| \sum_i \pi(i) P_{B_i}- \sum_i \pi(i) P_{B^*_i}\right\|_{\ell_p}\label{eq:convexity_2}\\
  &\leq & \sum_i \pi(i) \left\|  P_{B_i}-  P_{B^*_i}\right\|_{\ell_p}\label{eq:convexity_3}\\
&=&\sum_i \pi(i) D\left(P_{B_i}, P_{B^*_i}\right) \label{eq:convexity_4}\\
& = &\sum_i \pi(i) \mathcal{D}\left(P_{B_i}\right) \label{eq:convexity_5}
\end{eqnarray}
Eq. \eqref{eq:convexity_1} follows from the fact that $\sum_i \pi(i) P_{B^*_i}$ is a noncontextual behavior, 
Eqs. \eqref{eq:convexity_2}  and \eqref{eq:convexity_4} follow from the definition of  $D$,
Eq. \eqref{eq:convexity_3} follows from the convexity of the $\ell_p$ norm and \eqref{eq:convexity_5}
follows from the fact that each $B_i$ be the non-contextual behavior
achieving the minimum for $B_i^*$ in Eq. \eqref{eqdefdist}.
A similar argument shows that   $\mathcal{D}_u$ and $\mathcal{D}_{max}$ are also convex.

\subsubsection{Continuity}

Continuity of $\mathcal{D}$, $\mathcal{D}_u$ and $\mathcal{D}_{max}$ is a guaranteed by the continuity of the 
$\ell_p$ norms. This concludes the proof of Thm. \ref{teo_dist}.

 \hfill\qed\vspace{1ex}

\subsection{Proof of Theorem 5}
\label{sub:rob}

\setcounter{teo}{4}

\begin{teo}
 \begin{enumerate}
  \item The robustness of contextuality is faithful, convex and continuous;
  \item $\mathcal{R}\left(B_1 \& B_2\right) \leq \max_i \mathcal{R}\left(B_i\right)$;
  \item $\mathcal{R}\left(B_1 \otimes  B_2\right) \leq \mathcal{R}\left(B_1\right) + \mathcal{R}\left(B_2\right) - \mathcal{R}\left(B_1\right)\mathcal{R}\left(B_2\right)$;
 \item The contextual fraction can be calculated via linear programming.
 \end{enumerate}

\end{teo}

\noindent \textit{Proof.} In Ref. \cite{ABM17} the authors show that  the contextual fraction of a behavior $B$ is equal to $1- \boldsymbol{1}\cdot b^*$, where 
$b^ *$ is the subnormalized global probability distribution which is the optimal solution of the following linear program:

\begin{equation}
\begin{alignedat}{3}
&\textLP{Find }       \;\;&&  b \in \mathbb{R}^N
\\
&\textLP{maximising }   \;\;&& \boldsymbol{1} \cdot b
\\
&\textLP{subject to } \;\;&& M \, b \,\leq\, P_B
\\
&\textLP{and }        \;\;&& b \,\geq\, 0,
\end{alignedat}
\end{equation}
where $N$ is the number of contexts,  $\boldsymbol{1} \in \mathbb{R}^ N$ is the vectors with all entries equal to $1$ and $M$ is the \emph{incidence matrix} that records
the restriction relation between global assignments $g \in O^ {X}$ and  local assignments  $s \in O^ C$, that is,
\be M\left[s, g\right] = \begin{cases}
1 & \mbox{ if} \ \ g|_C=s;\\
0& \mbox{otherwise}. 
\end{cases} \ee

With a similar argument, one also proves that the robustness of $B$ is equal to $1- \frac{1}{\boldsymbol{1}\cdot b^*}$, where $b^*$ is the supernormalized probability distribution which is the  optimal solution of the 
following linear program:

\begin{equation}
\begin{alignedat}{3}
&\textLP{Find }       \;\;&&  b \in \mathbb{R}^N
\\
&\textLP{minimising }   \;\;&& \boldsymbol{1} \cdot b
\\
&\textLP{subject to } \;\;&& M \, b \,\geq\, P_B
\\
&\textLP{and }        \;\;&& b \,\geq\, 0,
\end{alignedat}
\end{equation}

The proof of Thm. \ref{teo:rob} follows exactly the same lines as the proof of Thm. \ref{teo:cf} presented in Ref. \cite{ABM17}.

 \hfill\qed\vspace{1ex}
 
 \subsection{Proof of Theorem 7}
 \label{approp}
 
 \setcounter{teo}{6}
 
 \begin{teo}
Given any graph $G=(V,E)$ we have that
$\Pi(\mathcal{Q}(G))\subset \mathcal{E}(G)$.
\end{teo}

\noindent \textit{Proof.}
Since both sets are convex, it is enough to prove that the extremal points of 
$\Pi(\mathcal{Q}(G))$ are contained in $\mathcal{E}(G)$. The extemal points of $\Pi(\mathcal{Q}(G))$ are obtained using pure state and hence if  $x \in \Pi(\mathcal{Q}(G))$ is a extremal point we have
\be x_{ij}=Tr(\ket{\psi}\bra{\psi}X_iX_j)=Tr(\ket{\psi}\bra{\psi}X_iX_j\ket{\psi}\bra{\psi})\ee
in which $X_i$ are quantum measurements with possible outcomes  $\pm 1$ with proper dimension. 

The set of matrices of the form $A=H\ket{\psi}\bra{\psi}$ where $H$ is hermitian is a real vector space
and \be\left\langle A, B\right\rangle = Tr\left(A^{\dagger}B\right)\ee
is an inner product in this vector space.  This means that there is an isomorphism between this set and some $\mathbb{R}^{k}$ that 
preserves the inner product. Each $\ket{\psi}\bra{\psi}X_i$ is connect with some  $u_i \in \mathbb{R}^{k}$ by this isomorphism 
and
\be x_{ij}=\braket{u_i}{u_j}.\ee
The vectors obtained this way are unitary, 
but may have more then $|V|$ coordinates. Since we have only $|V(G)|$ vectors, we can represent them in 
$\mathbb{R}^{|V(G)|}$ preserving the value of  $\braket{u_i}{u_j}$.

%This implies that
%\begin{eqnarray}
%x_{ij}&=&Tr(\ket{\psi}\bra{\psi}X_iX_j)\\
%&=&\sand{\psi}{X_iX_j}{\psi}\\
%&=&\braket{w_i}{w_j}
%\end{eqnarray}
%in which $w_i=X_i\ket{\psi}$. These vectors are unitary, since $X_i^2=I$, a property that follows from the fact that the
%eigenvectors of $X_i^2$ are  $\pm 1$.

%To conclude the proof, we have to show that we can replace the vectors $w_i$, which can be complex,
%by real vectors. To do that it is enough  to consider the vectors real vector $u_i$  obtained from 
%$w_i$ spliting real and imaginary parts of its coordinates.The vectors obtaied this way are unitary, 
%but may have more then $|V|$ coordinates. Since we have only $|V|$ vectors, we can represent them in 
%$\mathbb{R}^{|V|}$ preserving the value of  $\braket{u_i}{u_j}$.
\hfill\qed\vspace{1ex}

\subsection{Proof of Theorem 8}

\setcounter{teo}{7}

\begin{teo}
There is a point $z \in \mathcal{E}\left(C_n\right)$ for wich  
\be \sum_{i=0}^{n-2} z_{ii+1} - z_{0n-1} = n\cos\left(\frac{\pi}{n}\right).\ee
\end{teo}

To prove this fact, we first state some properties of the  $n$-cycle eliptope.
 
 \begin{prop}
 For any graph  $G $, the following are equivalent:
 \begin{enumerate}
 \item $\mathcal{E}(G) = \left\{ z \in [-1,1]^{|E(G)|} \left|  \frac{1}{\pi}\arccos(z) \in MET^{01}(G)\right.\right\};$
 \item $\mathcal{E}(G) = \left\{ z \in [-1,1]^{|E(G)|} \left| \frac{1}{\pi}\arccos(z) \in CUT^{01}(G)\right.\right\};$
 \item $G$ does not have any $K_4$ minor.
 \end{enumerate}
 \label{eliptopocut}
 \end{prop}
 
% %\textcolor{blue}{Referência: Complexity of the positive semidefinite matrix
% %completion problem with a rank constraint
% %M. E.-Nagy, M. Laurenty e A. Varvitsiotisz.
% %Geometry of Cuts and Metrics, pg 518 (teorema 31.3.7).}
% 
%  %\vspace{1em}
  See reference \cite{DL97} for a proof.
  Since no cycle has a  $K_4$ minor, we conclude that
 \begin{cor}
 For the  $n$-cycle $C_n$ we have
 \be \mathcal{E}(C_n) = \left\{ x \in [-1,1]^{|E|} \left| \frac{1}{\pi}\arccos(x) \in \mathrm{CUT}^{01}(C_n) \right.\right\}.\ee
 \end{cor}
 We can now  show that the eliptope of the $n$-cycle is larger than the quantum set. 
\vspace{1em}

\noindent \textit{Proof of Thm 8.} Such a point is
 \begin{eqnarray}
  z&=&\left(\cos\left(\frac{\pi}{n}\right),\cos\left(\frac{\pi}{n}\right), \ldots, -\cos\left(\frac{\pi}{n}\right)\right)\nonumber\\
  &=&\left(\cos\left(\frac{\pi}{n}\right),\cos\left(\frac{\pi}{n}\right), \ldots, \cos\left(\frac{(n-1)\pi}{n}\right)\right).\end{eqnarray}
 By Prop. \ref{eliptopocut}, to prove that 
 $z \in \mathcal{E}$ it is enough to prove that 
 \be y=\left(\frac{1}{n},\frac{1}{n}, \ldots, \frac{(n-1)}{n}\right) \in \mathrm{CUT}^{01}(C_n).\ee
  Since $\mathrm{CUT}^{01}$ and $\mathrm{CUT}^{\pm 1}$ are related by the map $\alpha$, $y \in \mathrm{CUT}^{01}(C_n) \Leftrightarrow \alpha(y) \in \mathrm{CUT}^{\pm 1}(C_n)$ and the last inclusion can be proven by showing that $\alpha(y)$ obeys all inequalities
\be \sum_{i=0}^{n-1} \gamma_{ii+1} \alpha(y)_{ii+1}\leq n-2\ee
 in which each coefficient $\gamma_{ii+1}=\pm 1$  and an odd naumber of them is equal to $-1$.  Since
\be \alpha(y)=\left(\frac{n-2}{n},\frac{n-2}{n}, \ldots, \frac{2-n}{n}\right)\ee
 and $\frac{n-2}{n} >0$ and $\frac{2-n}{n} < 0$, the largest value of $\sum_{i=0}^{n-1} \gamma_{ii+1} \alpha(y)_{ii+1}$ is $n-2$, 
 obtained when all coefficients $\gamma_{ii+1}$ are equal, except $\gamma_{0n-1}$. This implies that $\alpha(y) \in \mathrm{CUT}^{\pm 1}(C_n)
 \Rightarrow y \in \mathrm{CUT}^{01}(C_n) \Rightarrow   \in \mathcal{E}(C_n).$

 \hfill\qed\vspace{1ex}

%\subsection{Geometric distances for the $n$-cycle}

\subsection{A behavior violating more than one facet-defining inequality}
\label{app:3322}

For example, in the $(3,3,2,2)$  Bell scenario, the distribution 

\begin{table}[h!]
\begin{center}
  \begin{tabular}{c|c|c|c|c}
  &$11$&$1-1$&$-11$&$-1-1$\\
         $A_1B_1$&$\frac{1}{2}$&$0$&$0$&$\frac{1}{2}$  \\
        $A_1B_2$&$ \frac{1}{2}$&$0$&$0$&$\frac{1}{2}$   \\
         $A_1B_3$&$\frac{1}{2}$&$0$&$0$&$\frac{1}{2}$  \\
         $A_2B_1$&$\frac{1}{2}$&$0$&$0$&$\frac{1}{2}$ \\
        $A_2B_2$&$ 0$&$\frac{1}{2}$&$\frac{1}{2}$&$0$  \\
          $A_2B_3$ &$ 0$&$\frac{1}{2}$&$\frac{1}{2}$&$0$  \\
           $A_3B_1$& $\frac{1}{2}$&$0$&$0$&$\frac{1}{2}$\\
           $A_3B_2$& $\frac{1}{2}$&0&0 &$\frac{1}{2}$ \\
           $A_3B_3$ & $0$&$\frac{1}{2}$&$\frac{1}{2}$&$0$
        \end{tabular}
        \end{center}
        \end{table}
        where entrie of line $i$ and column $j$ represents probability of outcomes $j$ for measurement of context $i$,  
violates  the CHSH inequality and the $I_{3322}$ inequality, both facet-defining.

\subsection{An empirical model violating more than one facet-defining inequality}
\label{ap:2ineq}

For example, for the graph shown in Fig. \ref{fig1}, the noncontextuality inequalities
\begin{eqnarray*}
p_4+p_5+p_6+p_7+p_8 &\leq & 1, \\ 
2p_1+p_2+2p_3+2p_4+p_5+p_6+p_7+p_8 &\leq & 3
\end{eqnarray*}
are both facet defining and are both violated by the distribution that assigns $\frac{1}{3}$ to all vertices \cite{XC13}.